\pdfoutput=1
\documentclass[11pt,a4paper]{article}
\usepackage{jheppub} 

\usepackage[utf8]{inputenc}

\usepackage{amsfonts}
\usepackage{amssymb}
\usepackage{comment}
\usepackage{graphicx}
\usepackage{amsmath}
\usepackage{tabu}
\usepackage{dsfont}
\usepackage{float}
\usepackage{amsthm}
\usepackage{slashed}
\usepackage{setspace}
\usepackage[labelformat=simple]{subcaption}

\usepackage[font=small, labelfont=bf]{caption}
\usepackage{amsmath, amsthm, amssymb, amsfonts}
\usepackage{multirow}
\usepackage{graphicx}
\usepackage{float}
\usepackage[numbers,sort&compress]{natbib}
\usepackage{jheppub}
\usepackage{enumerate}

\def\bZ{\mathbb{Z}}

\newcommand{\nn}{\nonumber}

\newcommand{\p}{\partial}

\newcommand{\cN}{\mathcal{N}}

\newcommand{\cL}{\mathcal{L}}

\newcommand{\cO}{\mathcal{O}}
\newcommand{\cP}{\mathcal{P}}
\newcommand{\ee}{\mathrm{e}}

\newcommand{\I}{\mathrm{i}}
\newcommand{\im}{\mathrm{Im}}

\newcommand{\kom}{\, ,\quad }

\usepackage{comment}

\newcommand{\fft}[2]{\frac{#1}{#2}}
\newcommand{\ft}[2]{{\textstyle{\frac{#1}{#2}}}}

\newcommand{\nc}{\newcommand}
\nc{\beq}{\begin{equation}}
\nc{\eeq}{\end{equation}}
\nc{\bea}{\begin{eqnarray}}
\nc{\eea}{\end{eqnarray}}

\def\ov{\overline}

\usepackage{tikz} 
\usetikzlibrary{shapes,arrows,positioning,automata,backgrounds,calc,er,patterns}
\usepackage[compat=1.0.0]{tikz-feynman}

\usepackage{tikz-feynhand}

\allowdisplaybreaks[2]
\numberwithin{equation}{section}

\usepackage{footnotebackref}

\preprint{LITP-25-10, LMU-ASC 19/25}

\title{Type IIB at eight derivatives: Five-Point Axio-Dilaton Couplings}

\author[a]{James T. Liu,}
\author[b]{Ruben Minasian,}
\author[c]{Raffaele Savelli,}
\author[d,e]{Andreas Schachner}

\affiliation[a]{Leinweber Institute for Theoretical Physics, Randall Laboratory of Physics, University of Michigan,
Ann Arbor, MI 48109-1040, USA}
\affiliation[b]{Institut de Physique Th{\'e}orique, Universit{\'e} Paris Saclay, CNRS, CEA, 91191 Gif-sur-Yvette Cedex, France}
\affiliation[c]{Dipartimento di Fisica, Universit{\`a} di Roma Tor Vergata and INFN - Sezione di Roma Tor Vergata, via della Ricerca
Scientifica, I-00133 Roma, Italy}
\affiliation[d]{ASC for Theoretical Physics, LMU Munich, 80333 Munich, Germany}
\affiliation[e]{Department of Physics, Cornell University, Ithaca, NY 14853, USA}

\emailAdd{jimliu@umich.edu}
\emailAdd{ruben.minasian@ipht.fr}
\emailAdd{raffaele.savelli@roma2.infn.it}
\emailAdd{a.schachner@lmu.de}
\emailAdd{as3475@cornell.edu}

\abstract{

We study the Type IIB eight-derivative effective Lagrangian beyond the quartic level, focusing on interactions involving gravitons and axio-dilatons. We show how to translate five-point scattering amplitudes into genuine five-point contact terms and extract all perturbative contributions to the effective action. Our result is consistent with T-duality predictions at tree level in the NSNS sector. We find that couplings with an even/odd number of scalars are neutral/charged under $\mathrm{SL}(2,\mathbb{Z})$, and use this feature to deduce their non-perturbative completion. The mixed NSNS/RR structures that cannot be deduced from the pure NSNS sector allow us to unambiguously fix the kinematics, which turns out to be the same for tree level and one loop. In Type IIA, these structures are essential for establishing agreement of string-theoretic corrections with the circle reduction of M-theory higher-derivative couplings.

}

\keywords{}

\begin{document}

\maketitle

\bigskip

\newpage

\section{Introduction}

Despite decades of study of higher-derivative corrections to the ten-dimensional effective actions of string theory \cite{Gross:1986mw, Gross:1986iv}, their complete structure remains elusive. These corrections play a crucial role in shaping our understanding of low-energy string dynamics, as they govern the form of perturbative $\alpha'$ corrections that appear upon compactification. While recent advances in string-amplitude techniques have significantly improved computational reach, no unifying framework has emerged for constructing the full effective action at higher orders in $\alpha'$ and $g_{s}$. A central challenge is the construction of effective actions reproducing string amplitudes on-shell, which is complicated by the proliferation of kinematical structures, subtle pole-subtraction procedures, and field redefinitions.

In the NSNS sector, partial results for quintic and sextic terms have been obtained in \cite{Peeters:2001ub, Liu:2013dna, Liu:2019ses}. For Type IIB, the quartic couplings involving RR fields were completed in \cite{Policastro:2006vt, Policastro:2008hg}, while the supersymmetric completion of $R^4$ with $F_5$ was inferred via $\mathcal{N}=2$ superspace techniques \cite{Howe:1983sra, deHaro:2002vk, Green:2003an, Rajaraman:2005ag, Paulos:2008tn}, with applications to D3-brane backgrounds and AdS$_5 \times S^5$ black holes \cite{Green:2003an, Paulos:2008tn, Melo:2020amq}. A complete tree-level NSNS-sector action was proposed in \cite{Garousi:2020gio,Garousi:2020lof} using T-duality constraints.  (See also \cite{Wulff:2021fhr,Wulff:2024mgu}.)  More recently, combining superstring amplitude techniques with Schwinger-type computations in M-theory and superfield methods, we shed new light on the structure of the effective action for the complexified three-form $G_{3}$ at the five-point level \cite{Liu:2022bfg}.

In this work, we study the structure of the Type IIB eight-derivative effective action beyond the quartic level by computing five-point scattering amplitudes involving the dilaton, the RR axion, and the graviton. Using the pure spinor formalism and a detailed pole-subtraction procedure, we isolate genuine five-point contact terms and determine their precise kinematical structure at both tree level and one loop. This allows us to write down the full $\mathrm{SL}(2,\mathbb{Z})$-invariant effective action at this order in Sec.~\ref{sec:FinalEFT}.

Earlier attempts to constrain higher-derivative couplings using $\mathrm{SL}(2,\mathbb{Z})$ modular invariance such as \cite{Kehagias:1997cq} were limited by ambiguities in the modular completion of the terms. Following the approach of \cite{Policastro:2008hg}, we resolve these ambiguities by explicitly computing five-point mixed-sector amplitudes involving both NSNS and RR states. In addition, we exploit a key supersymmetry constraint: The maximal allowed violation of $\mathrm{U}(1)$ R-symmetry in a $P$-point amplitude is bounded by $2(P - 4)$ \cite{Boels:2012zr,Green:2019rhz}. This also provides a direct link to the full $\mathrm{SL}(2,\mathbb{Z})$ modular completion of the effective action. For five-point interactions involving dilatons $\phi$, universal axions $\chi$, and gravitons $h$, we have to distinguish two distinct categories.
\begin{enumerate}
	\item \emph{Even number of scalars}: The $\mathrm{U}(1)$ charge vanishes, and the modular completion is uniquely fixed by the non-holomorphic Eisenstein series $f_0(\tau,\bar{\tau})$ of weight $3/2$ which also governs the $R^4$ couplings \cite{Green:1997as}.
	\item \emph{Odd number of scalars}: The couplings are maximally $\mathrm{U}(1)$-violating (MUV), and as we demonstrate explicitly, their kinematical structure is fixed by a unique index contraction. Their modular completion is controlled by the modular forms $f_{\pm1}(\tau, \bar{\tau})$, in agreement with expectations from duality to M-theory.
\end{enumerate}

One of the key results of the paper is the explicit determination of the five-point, eight-derivative effective action in the scalar-graviton sector of Type IIB. We confirm that our results are consistent with earlier findings at tree level in the pure NSNS sector obtained through T-duality \cite{Garousi:2020gio,Garousi:2020lof}. Along the way we clarify some confusions in the literature, related to the presence of the dilaton in the Type II string-frame eight-derivative effective action.  While it is possible to pick a field basis where the NSNS string-frame Lagrangian has no explicit dilaton dependence, this may no longer be true once couplings to the RR sector are taken into account.  Nevertheless, at quartic order, all dilaton couplings in the string frame can be replaced by corresponding Ricci terms through a suitable field redefinition.  For this reason, the presence or absence of dilaton couplings in the string frame in itself has no direct effect on physical string amplitudes.  In Type IIA string theory at one loop, the role of the dilaton in the string-frame $(\alpha')^3$ effective action is correctly captured by the circle reduction of M-theory eight-derivative terms, such as $R^4$.

The paper is organised as follows. We start, in Sec.~\ref{sec:review}, by setting the stage of our investigation and by describing the strategy we adopt to carry it out. In Sec.~\ref{sec:NSNSdilaton}, we treat the pure NSNS sector and translate five-point dilaton/graviton scattering amplitudes at tree level and one loop into genuine five-point couplings in the eight-derivative effective Lagrangian. We comment on their structure and verify their compatibility with T-duality. In Sec.~\ref{sec:RRcomp}, we extend the analysis to the RR sector and fix the subtle kinematics governing the mixed NSNS/RR couplings. We then use the power of modular invariance to write down the full effective action, exact in the string coupling. Sec.~\ref{sec:IIAMth} takes a closer look at the nature of dilaton couplings in the string-frame action and zooms in on the case of Type IIA at eight derivatives and its uplift to M-theory. We conclude in Sec.~\ref{sec:Concl} with a few suggestions for future studies. App.~\ref{app:strings} contains a detailed account of the pole-subtraction procedure which is at the core of our methodology, and App.~\ref{app:Ftheory} elaborates more on an interesting hint for further investigation, namely the question of whether the Type IIB action beyond leading order in $\alpha'$ admits a twelve-dimensional formulation.

\section{Context and methodology}\label{sec:review}

In this section, we set the stage of our investigation by reviewing, in Sec.~\ref{Sec:4points}, the well-known eight-derivative Lagrangian of Type IIB string theory at four points. Then, we will proceed by explaining, in Sec.~\ref{Sec:Method}, the strategy we will follow to extract the effective couplings at five points. We will keep the discussion general and defer all technical details to App.~\ref{app:strings}.

\subsection{The four-point effective action}\label{Sec:4points}

Type IIB supergravity is the low-energy effective field theory that describes the dynamics of massless modes in Type IIB string theory. It is a chiral 10-dimensional theory with maximal $\mathcal{N} = 2$ supersymmetry, whose bosonic field content includes the metric $g_{\mu\nu}$, a complex scalar called the axio-dilaton $\tau$, a complex three-form $G_{3}$, and a self-dual five-form field strength. In particular, we use the convention where
\begin{equation}\label{eq:tauPG}
\tau = \chi+\I\,\mathrm{e}^{-\phi}\kom \cP_{\mu}=\dfrac{\I\,\mathrm{e}^{\phi}}{2}\nabla_{\mu}\tau\kom G_3 = \mathrm{e}^{\phi/2}\bigl (F_3-\tau H_3\bigl )\, .
\end{equation}

At leading order in the $\alpha'$ expansion, the effective action of Type IIB supergravity contains two-derivative terms which are invariant under a global $\mathrm{SL}(2,\mathbb{R})$ symmetry and a local $\mathrm{U}(1)$ R-symmetry with $G_{3}$ and $\cP$ carrying $+1$ and $+2$ units of R-charge respectively.
Beyond this two-derivative approximation, stringy corrections arise in the form of higher-derivative terms in the effective action, organised as an expansion in powers of $\alpha'$. The most well-known such correction is the eight-derivative $\mathcal{O}(\alpha'^3)$ term of the schematic form $R^4$, which arises from scattering amplitudes at tree level and one loop \cite{Green:1981ya,Gross:1986iv}, but also receives contributions from an infinite tower of D-instantons \cite{Green:1997tv,Green:1997as}. 

To identify effective couplings beyond string perturbation theory, S-duality plays a key role. Indeed, after quantisation, only a $\mathrm{SL}(2,\mathbb{Z})$ subgroup of $\mathrm{SL}(2,\mathbb{R})$ survives, the Type IIB duality group. Moreover, configurations connected by duality are to be regarded as physically indistinguishable. The axio-dilaton transforms under $\mathrm{SL}(2,\bZ)$ according to
\begin{equation}
\tau\rightarrow \dfrac{a\tau+b}{c\tau+d}\kom ad-bc=1\, ,
\end{equation}
and therefore the fields $\cP$ and $G_3$ pick up just a phase:
\begin{equation}
\cP\rightarrow \dfrac{c\bar{\tau}+d}{c\tau+d}\, \cP\kom G_{3}\rightarrow \left (\dfrac{c\bar{\tau}+d}{c\tau+d}\right )^{\frac{1}{2}}G_{3}\, .
\end{equation}
Composite terms $\Phi$ corresponding to a combination of fields carrying a total $\mathrm{U}(1)$-charge $2w$ transform with weight $w$, so that
\begin{equation}
\Phi\rightarrow\left ( \dfrac{c\bar{\tau}+d}{c\tau+d}\right )^{w} \Phi\, .
\end{equation}
Invariance of the effective action under $\mathrm{SL}(2,\mathbb{Z})$ requires that each contact term individually respects this symmetry. At the higher-derivative level, this significantly restricts the allowed coefficients as functions of $\tau$. In particular, they must take the form of $\mathrm{SL}(2,\mathbb{Z})$-covariant modular forms.

Following this general strategy, the complete quartic effective action has been derived and takes the following form \cite{Policastro:2006vt,Policastro:2008hg, Liu:2019ses}\footnote{We include for completeness the structure $\epsilon_{8}\epsilon_{8}$, although kinematics forces it to start contributing non-trivially at five points.}
\begin{align}
\cL_{\text{4-pt}}^{(3)}&=\alpha\, f_{0}(\tau,\bar{\tau})\biggl \{\left ( t_{8}t_{8}-\tfrac{1}{4}\epsilon_{8}\epsilon_{8}\right )\biggl [R^4+6R^{2}\left (4|\nabla\cP|^{2}+|\nabla G_{3}|^{2}\right )+24|\nabla \cP|^{2}|\nabla G_{3}|^{2}\nn\\[0.4em]
&\quad+12R\left (\nabla\cP(\nabla\overline{G}_{3})^{2}+\nabla\overline{\cP}(\nabla{G}_{3})^{2}\right )\biggl ]+\cO_{1}\left ((|\nabla\cP|^{2})^{2}\right )+\cO_{2}\left ((|\nabla G_{3}|^{2})^{2}\right )
\biggl\}
\label{eq:PTaction}
\end{align}
where we introduced
\begin{equation}\label{eq:DefAlpha} 
 \alpha=\dfrac{(\alpha^{\prime})^{3}}{3\cdot 2^{12}}\, , \; (\nabla\cP)_{\mu\nu}\,^{\rho\sigma}=\delta_{[\mu}^{[\rho}\nabla_{\nu]} \cP^{\sigma]}\, .
\end{equation}
The four-point kinematics is mostly specified in terms of the $t_8$ tensor \cite{Schwarz:1982jn} (see Eq.~\eqref{eq:t8} in App.~\ref{app:strings} for the definition) and of the two operators $\cO_{1}$ and $\cO_{2}$ accounting for the different kinematical structure of terms mixing NSNS and RR sector (see e.g.~\cite{Liu:2019ses}). In what follows, we will only be interested in the first one, which, based on \cite{Minasian:2015bxa}, we take of the form
\begin{equation}
\label{eq:O1}
    \mathcal O_1\left((|\nabla\cP|^{2})^{2}\right)=1536(\nabla_{\mu}\cP_{\nu}\nabla^{\mu}\cP^{\nu})(\nabla_{\sigma}\overline \cP_{\rho}\nabla^{\sigma}\overline{\cP}^{\rho})\, .
\end{equation}
Finally, the prefactor $f_0(\tau,\bar{\tau})$ in Eq.~\eqref{eq:PTaction} corresponds to the non-holomorphic Eisenstein series of weight $3/2$ (see e.g.~\cite{Green:2019rhz})
\begin{equation}\label{eq:f0}
    f_0(\tau,\bar{\tau}) = \sum_{(\ell_{1},\ell_{2})\neq (0,0)}\, \frac{\im(\tau)^{3/2}}{\bigl|\ell_{1}+\tau\ell_{2}\bigl|^{3}}\, .
\end{equation}

\subsection{Beyond four points}\label{Sec:Method}

Beyond four points, an important new phenomenon appears in the effective Lagrangian, namely the presence of kinematical structures which are not invariant under $\mathrm{U}(1)$. In fact, the number of points $P$, or equivalently the number of external states in the amplitude, is related to the maximal $\mathrm{U}(1)$ charge $Q_{\rm max}$ of the corresponding coupling by $|Q_{\rm max}|=2(P-4)$ \cite{Green:1997me}.\footnote{This implies e.g.~that terms of the schematic form $\cP^{2}R^{3}$ are forbidden, as can also be confirmed by studying $\cN=2$ compactifications to $4$D (see Sec.~\ref{sec:evenD}).}
More recently, special kinematical structures in the so-called maximally U$(1)$-violating (MUV) amplitudes were observed in \cite{Boels:2012zr} and further investigated in \cite{Green:2019rhz}, which naturally arise from superspace integrals \cite{Howe:1983sra,Peeters:2001ub,deHaro:2002vk,Green:2003an,Green:2005qr,Peeters:2005tb} and do not factorise into combinations of $\epsilon_{n}$ or $t_{8}$.

In \cite{Liu:2022bfg} we have derived part of the five-point Type IIB effective action, which indeed features the aforementioned behaviour. There we concentrated on the gravity plus $G_3$-flux sector, where e.g.~we uncovered the following MUV structure
\begin{align}\label{eq:FullResultG2R3TreeLoopxx} 
\cL_{G_3R}^{(3)}&\supset \frac{3}{2}\alpha \left (f_{1}(\tau,\bar{\tau})\, t_{18}G^{2}_{3}R^{3}+f_{-1}(\tau,\bar{\tau})\, t_{18}\ov G^{2}_{3}R^{3}\right )\,,
\end{align}
where $t_{18}$ is one such higher tensor structure and $f_1(\tau,\bar\tau)$ is a non-holomorphic function of $\tau$ whose $\mathrm{SL}(2,\mathbb{Z})$ transformation compensates that of $G_3^2R^3$, making the Lagrangian invariant. In general, the charged generalisations of \eqref{eq:f0} are defined as
\begin{equation}\label{eq:ModFormsDef} 
    f_{w}(\tau,\bar{\tau})=\sum_{(\ell_{1},\ell_{2})\neq (0,0)}\, \dfrac{\im(\tau)^{3/2}}{(\ell_{1}+\tau\ell_{2})^{\frac{3}{2}+w}(\ell_{1}+\bar{\tau}\ell_{2})^{\frac{3}{2}-w}}\, ,
\end{equation}
which transform under $\mathrm{SL}(2,\mathbb{Z})$ as
\begin{equation}
    f_{w}\left (\dfrac{a\tau+b}{c\tau+d},\dfrac{a\bar{\tau}+b}{c\bar{\tau}+d}\right )=\left (\dfrac{c\tau+d}{c\bar{\tau}+d}\right )^{w}f_{w}(\tau,\bar{\tau})\, .
\end{equation}
By expanding $f_{w}$ at weak string coupling, i.e.~for $\mathrm{Im}(\tau)\gg 1$, one finds
\begin{equation}\label{eq:ExpansionModFuncLargeImTauK} 
    f_{w}(\tau,\bar{\tau})=a_{T}+\dfrac{a_{L}}{(1-4w^{2})}+\cO\left (\ee^{-\mathrm{Im}(\tau)}\right )
\end{equation}
in terms of the closed-string tree-level coefficient $a_T$ \cite{Gross:1986iv} and one-loop coefficient $a_L$ \cite{Green:1981ya}:
\begin{equation}\label{eq:TLCMFEXP} 
    a_{T}=2\zeta(3)\mathrm{Im}(\tau)^{\frac{3}{2}}\kom a_{L}=\dfrac{2\pi^{2}}{3}\mathrm{Im}(\tau)^{-\frac{1}{2}}\, .
\end{equation}
The exponentially suppressed terms encode non-perturbative contributions from D$(-1)$-instantons \cite{Green:1997tv}.
Therefore, up to five-point interactions, the only modular functions that appear are\footnote{From now on, the coefficients $a_T$ and $a_L$ will be understood to multiply tree-level and one-loop terms respectively, and will no longer be written explicitly.}
\begin{align}\label{eq:fpm1}
f_{0}(\tau,\bar{\tau})&=a_{T}+a_{L}+\cO\bigl (\mathrm{e}^{-\mathrm{Im}(\tau)}\bigl )\kom f_{\pm 1}(\tau,\bar{\tau})=a_{T}-\tfrac{1}{3}a_{L}+\cO\bigl (\mathrm{e}^{-\mathrm{Im}(\tau)}\bigl )\, .
\end{align}

However, extracting five-point couplings from the corresponding five-point scattering amplitudes is not as straightforward as for the four-point case. This is essentially due to two reasons:\footnote{Clearly, these reasons are also present when going from four-point amplitudes to the quartic effective action. However, the very limited number of terms in the two-derivative action and, consequently, of exchange channels, makes this process way more straightforward than the analogous one at five points.}
\begin{itemize}
\item The five-point amplitudes may factorise into lower-point amplitudes. This structure typically appears as poles\footnote{Sometimes it can happen that one of the vertices has a momentum dependence that cancels the one coming from the propagator of the exchanged particle, resulting in a finite contribution.} in the amplitude, corresponding to exchange of intermediate massless states.
\item The five-point amplitudes may contain information about the non-linear completion of lower-point couplings. For instance, the quantity $\nabla\cP$ appearing in Eq.~\eqref{eq:PTaction} contributes two terms
\begin{equation}
    (\nabla\cP)_{\mu\nu}\,^{\rho\sigma}=\dfrac{\I}{2\tau_2}\delta_{[\mu}^{[\rho}\nabla_{\nu]}\nabla^{\sigma]}\tau-\dfrac{\I}{2\tau_2^2}\delta_{[\mu}^{[\rho}\nabla_{\nu]}\tau_2\nabla^{\sigma]}\tau\,,
\end{equation}
where the second term is the non-linear completion of the first one, as it is quadratic in the derivatives of the scalars%
\footnote{Non-linear couplings to the metric fluctuations $h_{\mu\nu}$ also appear in $\nabla\mathcal P$, both from the raised spacetime indices and from the Christoffel connections implicit in the covariant derivatives.}.
\end{itemize}
In order to identify the genuine five-point contact terms, one needs to suitably remove all contributions coming from both of the mechanisms described above. Once the factorised five-point function has been removed from a given five-point amplitude, the five-point contact interaction that is left need not necessarily be a genuine one yet. Only after removing also the part that comes from the non-linear completion of lower-point contact terms, does one extract the genuinely-five-point couplings. 

In App.~\ref{app:strings}, we provide a detailed account of this procedure, which we will heavily use to derive the new five-point couplings and consequently the effective action in Sec.~\ref{sec:NSNSdilaton} and Sec.~\ref{sec:RRcomp}.

\section{Five-point contact terms in the NSNS sector}\label{sec:NSNSdilaton} 

Our goal is to construct the complete quintic effective action of the Type IIB string in the gravity and scalar sector.  This will be done by analysing five-point scattering amplitudes involving gravitons, dilatons and the RR axion.  We first focus on the NSNS sector, because the tensor structure of the couplings can be accessed more directly. In the next section, we will extend our analysis to mixed sectors involving RR fields.

We begin in Sec.~\ref{sec:AmpRes} by outlining the origin of the relevant five-point amplitudes both at tree level and at one loop, and then proceed to isolate the genuine higher-derivative contributions, as described in Sec.~\ref{Sec:Method}. In Sec.~\ref{sec:Tduality}, we prove that our proposal for the five-point effective action at tree level in the NSNS sector satisfies the constraints imposed by T-duality. Finally, in Sec.~\ref{sec:evenD} and Sec.~\ref{sec:oddD}, we zoom into the couplings containing respectively an even and an odd number of dilatons, and provide a more detailed description of their structure and features.

\subsection{Tree and loop dilaton-graviton amplitudes at five points}\label{sec:AmpRes} 

The scattering of massless closed-string NSNS states is treated uniformly, with the only distinction being the choice of symmetric, antisymmetric or trace polarisations.  Thus there is only a single tree-level five-point amplitude, \eqref{eq:M5t}, and a single one-loop five-point amplitude, \eqref{eq:M5l}, to deal with.  By substituting in the appropriate polarisations, and performing the subtractions described in Sec.~\ref{Sec:Method}, we can obtain the corresponding five-point contact terms corresponding to the $h^{5}$, $h^{4}\phi$, $h^{3}\phi^{2}$, $h^{2}\phi^{3}$, $h\phi^{4}$, and $\phi^{5}$ amplitudes.

The first step in obtaining the quintic effective action is to subtract the pole terms that arise from factorisation on the four-point amplitude.  Since the four-point amplitude is identical at tree and loop level, the subtraction procedure is the same in both cases.  Note that, since the effective action can be shifted by field redefinitions, it is important to be precise about the quartic action that we are using.  Here, we use the gravity/scalar sector of (\ref{eq:PTaction}), which at the four-point function level reduces to%
\footnote{Note that this differs from the natural choice $\cL_{4-\mathrm{pt}}=\alpha t_8t_8(R^4+6R^2(\nabla\nabla\phi)^2+(\nabla\nabla\phi)^4)$ motivated by the Weyl-scaled Riemann tensor, $\bar R^{\mu\nu}{}_{\rho\sigma}=R^{\mu\nu}{}_{\rho\sigma}-\delta^{[\mu}{}_{[\rho}\nabla^{\nu]}\nabla_{\sigma]}\phi$.  However, for the IIB theory, it is appropriate to use the SL(2,$\mathbb{Z})$-invariant form of the $\mathcal O_1$ operator in (\ref{eq:PTaction}), which is what leads to the last term in (\ref{eq:NSNS4}).}
\begin{equation}
    \cL_{\text{4-pt}}=\alpha\Bigl(t_8t_8\left(R^{4}+6R^{2}(\nabla \nabla \phi)^{2}\right)+96((\nabla_{\mu}\nabla_{\nu}\phi)^{2})^2\Bigr)\,,
\label{eq:NSNS4}
\end{equation}
in the NSNS sector.  After subtracting these underlying four-point terms, we find that the resulting five-point contact terms can be reproduced by the effective couplings shown in Table~\ref{tab:FivePointFunction}.

\begin{table}[t]
\centering
\resizebox{\columnwidth}{!}{%
\begin{tabular}{|c||c|c|}
\hline 
&&\\[-0.8em]
 & Tree & Loop \\ [0.4em]
 \hline
\hline 
& \multicolumn{2}{c|}{} \\[-0.8em]
$h^{5}$ &\multicolumn{2}{c|}{ $\bigl (t_{8}t_{8}-\frac{1}{4}\epsilon_{8}\epsilon_{8}\bigl )R^{4}|_{\rm 5\, pts}$ }\\[0.4em]
\hline 
&&\\[-0.8em]
$\phi h^{4}$ & $-\frac{3\phi}{2}\bigl (t_{8}t_{8}-\frac{1}{4}\epsilon_{8}\epsilon_{8}\bigl )R^{4}$ & $\frac{\phi}{2}\bigl (t_{8}t_{8}-\frac{1}{4}\epsilon_{8}\epsilon_{8}\bigl )R^{4}$ \\ [0.4em]
\hline 
& \multicolumn{2}{c|}{} \\[-0.8em]
$\phi^{2} h^{3}$ & \multicolumn{2}{c|}{$\bigl (t_{8}t_{8}-\frac{1}{4}\epsilon_{8}\epsilon_{8}\bigl )\bigl (6R^{2}(\nabla \nabla \phi)^{2}|_{\rm 5\, pts}+4R^{3}\bigl [\frac{3}{4}\delta^{\nu}_{\mu} (\p_{\sigma}\phi)(\p^{\lambda}\phi)-\frac{1}{24}\delta^{\nu}_{\mu}\delta^{\lambda}_{\sigma} (\p \phi)^{2}\bigl ]\bigl )$ }   \\ [0.4em]
\hline 
&&\\[-0.8em]
$\phi^{3} h^{2}$ & $-\frac{3\phi}{2}\bigl (t_{8}t_{8}-\frac{1}{36}\epsilon_{8}\epsilon_{8}\bigl )\, 6R^{2}(\nabla\nabla \phi)^{2}$ & $\frac{\phi}{2}\bigl (t_{8}t_{8}-\frac{1}{36}\epsilon_{8}\epsilon_{8}\bigl )\, 6R^{2}(\nabla\nabla \phi)^{2}$ \\ [0.4em]
\hline 
& \multicolumn{2}{c|}{} \\[-0.8em]
$\phi^{4} h$ & \multicolumn{2}{c|}{$96((\nabla_{\mu}\nabla_{\nu}\phi)^2)^2|_{\rm 5\, pts}+24R^{\mu\nu\rho\sigma}(\nabla_\mu\nabla_\rho\phi\nabla_\nu\nabla_\sigma\phi (\partial\phi)^2-4\nabla_\mu\nabla_\alpha\phi\nabla_\rho\nabla^\alpha\phi\,\partial_\nu\phi\partial_\sigma\phi)$ }   \\ [0.4em]
\hline 
&&\\[-0.8em]
$\phi^{5}$ & $-\fft{3\phi}296((\nabla_{\mu}\nabla_{\nu}\phi)^2)^2+\fft4{75}\phi\epsilon_8\epsilon_8(\nabla\nabla\phi)^4$ & $\frac{\phi}{2}96((\nabla_{\mu}\nabla_{\nu}\phi)^2)^2-\fft4{225}\phi\epsilon_8\epsilon_8(\nabla\nabla\phi)^4$ \\ [0.4em]
\hline 
\end{tabular}%
}
\caption{Five-point contact terms, obtained by pole subtraction from the relevant five-point amplitudes. The even-dilaton couplings are not divided into tree and loop, because they only differ for the coefficient, $a_T$ vs $a_L$ in Eq.~\eqref{eq:TLCMFEXP}.}\label{tab:FivePointFunction} 
\end{table}

It is worth noting that the contact terms given in Table~\ref{tab:FivePointFunction} are not necessarily unique, as they can be shifted by field redefinitions.  The choices made here are motivated by demanding that the terms be written as $t_8t_8$ and $\epsilon_8\epsilon_8$ combinations, except in the case of $\phi^4h$, where this was not possible.  We will have more to say about such field redefinitions below when discussing the structure of the various couplings.

While the terms in Table~\ref{tab:FivePointFunction} directly correspond to the contact terms in the five-point string amplitudes, we are not yet done in extracting the effective action.  As mentioned in Sec.~\ref{Sec:Method}, we must still subtract the non-linear components of the quartic effective action, (\ref{eq:PTaction}), in order to obtain the true five-point couplings.  Some of these terms are explicitly shown in Table~\ref{tab:FivePointFunction}; the label ``5 pts'' indicates the presence of a graviton in the five-point amplitude.  In addition, for the odd-dilaton amplitudes, we have to account for the expansion of the exponential, $\ee^{n\phi}$, in the amplitudes
\begin{align}
\label{eq:tree4pt}
\text{tree:}\quad&\ee^{-\frac{3}{2}\phi}\Bigl(\bigl (t_{8}t_{8}-\frac{1}{4}\epsilon_{8}\epsilon_{8}\bigl )\left [R^{4}+6R^{2}(\nabla \nabla \phi)^{2}\right]+96((\nabla_{\mu}\nabla_{\nu}\phi)^{2})^2\Bigr)\nn\\
&\longrightarrow\quad-\fft32\phi\Bigl(\bigl (t_{8}t_{8}-\frac{1}{4}\epsilon_{8}\epsilon_{8}\bigl )\left [R^{4}+6R^{2}(\nabla \nabla \phi)^{2}\right]+96((\nabla_{\mu}\nabla_{\nu}\phi)^{2})^2\Bigr)\,,
\end{align}
and
\begin{align}
\text{loop:}\quad& \ee^{\frac{1}{2}\phi}\Bigl(\bigl (t_{8}t_{8}-\frac{1}{4}\epsilon_{8}\epsilon_{8}\bigl )\left [R^{4}+6R^{2}(\nabla \nabla \phi)^{2}\right]+96((\nabla_{\mu}\nabla_{\nu}\phi)^2)^2\Bigr)\nn\\
&\longrightarrow\quad\fft12\phi\Bigl(\bigl (t_{8}t_{8}-\frac{1}{4}\epsilon_{8}\epsilon_{8}\bigl )\left [R^{4}+6R^{2}(\nabla \nabla \phi)^{2}\right]+96((\nabla_{\mu}\nabla_{\nu}\phi)^{2})^2\Bigr)\,.
\end{align}
After the non-linear completion of the quartic terms is removed, we are left with the genuine five-point contact terms listed in Tab.~\ref{tab:FivePointContact}.

\begin{table}[t]
\centering
\begin{tabular}{|c||c|c|}
\hline 
&&\\[-0.8em]
 & Tree & Loop \\ [0.4em]
 \hline
\hline 
& \multicolumn{2}{c|}{}\\[-0.8em]
$h^{5}$ &  \multicolumn{2}{c|}{ $0$ }\\[0.4em]
\hline 
&&\\[-0.8em]
$\phi h^{4}$ & $0$ & $0$ \\ [0.4em]
\hline 
& \multicolumn{2}{c|}{} \\[-0.8em]
$\phi^{2} h^{3}$ & \multicolumn{2}{c|}{$\bigl (t_{8}t_{8}-\frac{1}{4}\epsilon_{8}\epsilon_{8}\bigl )\, 4R^{3}\bigl [\frac{3}{4}\delta^{\nu}_{\mu} (\p_{\sigma}\phi)(\p^{\lambda}\phi)-\frac{1}{24}\delta^{\nu}_{\mu}\delta^{\lambda}_{\sigma} (\p \phi)^{2}\bigl ]$ }   \\ [0.4em]
\hline 
&&\\[-0.8em]
$\phi^{3} h^{2}$ & \hspace*{0.75cm} $2 \epsilon_{8}\epsilon_{8}\,  R^{2}(\p\phi)^{2}(\nabla\nabla \phi)$  \hspace*{0.75cm} & $-\frac{2}{3} \epsilon_{8}\epsilon_{8}\,  R^{2}(\p\phi)^{2}(\nabla\nabla \phi)$ \\ [0.4em]
\hline 
& \multicolumn{2}{c|}{} \\[-0.8em]
$\phi^{4} h$ & \multicolumn{2}{c|}{\hspace*{0.625cm}$24R^{\mu\nu\rho\sigma}(\nabla_\mu\nabla_\rho\phi\nabla_\nu\nabla_\sigma\phi \,(\partial\phi)^2-4\nabla_\mu\nabla_\alpha\phi\nabla_\rho\nabla^\alpha\phi\,\partial_\nu\phi\partial_\sigma\phi)$\hspace*{0.625cm} }   \\ [0.4em]
\hline 
&&\\[-0.8em]
$\phi^{5}$ & $-\frac{4}{75} \epsilon_{8}\epsilon_{8}\,  (\p \phi)^{2}(\nabla\nabla \phi)^{3}$ & $\frac{4}{225} \epsilon_{8}\epsilon_{8}\,  (\p \phi)^{2}(\nabla\nabla \phi)^{3}$ \\ [0.4em]

\hline 
\end{tabular} 
\caption{Genuine five-point contact terms, obtained from those of Table~\ref{tab:FivePointFunction} by subtracting the part due to the non-linear completion of the four-point couplings.}\label{tab:FivePointContact} 
\end{table}

\subsection{A tree-level match with T-duality}\label{sec:Tduality}

In the NSNS sector, the full set of eight-derivative tree-level couplings has been derived in \cite{Garousi:2020gio,Garousi:2020lof} based on T-duality invariance.  In the absence of the antisymmetric tensor, the result is particularly simple in the string frame
\begin{equation}
    \cL=384\alpha e^{-2\phi}(2R_{\mu\nu\rho\sigma}R^{\mu\lambda\rho\delta}R_{\lambda\zeta}{}^{\nu\eta}R_{\delta\eta}{}^{\sigma\zeta}+R_{\mu\nu}{}^{\rho\sigma}R^{\mu\nu\gamma\delta}R_{\gamma\zeta\rho\eta}R_\delta{}^\eta{}_\sigma{}^\zeta).
\label{eq:garousi}
\end{equation}
This matches the sigma-model \cite{Grisaru:1986vi,Freeman:1986zh} and string \cite{Gross:1986iv} computations, as expected, except that here there are no Ricci terms at all.  This result indicates that the effective action can be written in a field-redefinition frame where the dilaton is completely absent in the string frame.

Since the action constructed by T-duality is unique \cite{Garousi:2020gio}, it should match the string amplitude calculation.  In order to verify this match, we first Weyl-scale (\ref{eq:garousi}) to the Einstein frame.  It is easy to see that the Einstein frame action will pick up a large number of dilaton couplings, extending all the way to the eight-point coupling $((\partial\phi)^2)^4$.   What is more noteworthy for comparison to the string five-point amplitude, however, is that the Einstein-frame action will pick up Ricci terms that result from the contraction of Riemann tensors with metric factors that arise from the Weyl scaling.  Such Ricci terms are not present in the effective contact terms given in Table~\ref{tab:FivePointContact}.  However, they can be removed by field redefinitions.  In practice, this means we may make use of the equations of motion to substitute
\begin{equation}
    R_{\mu\nu}\to\fft12\partial_\mu\phi\partial_\nu\phi,\qquad\Box\phi\to0.
\end{equation}
By making use of these field redefinitions, as well as integration by parts and Riemann identities, we have verified that the quintic effective action indeed matches (\ref{eq:garousi}).

At five points, the aforementioned matching of tree-level couplings is sufficient to uniquely determine the one-loop contributions as well. This follows from the observed factorisation of the kinematic structures, as summarised in Table~\ref{tab:FivePointContact}, and further discussed in the following subsections. A potential implication of this matching is that, while in string frame the tree-level effective action can be written without explicit dilaton terms, this may not simultaneously be possible at one loop. The reason lies in the fact that the tree-level and one-loop amplitudes share the same kinematical structures but differ in their relative coefficients. As a result, one would assume that not all dilaton couplings can be eliminated simultaneously through a single choice of field basis in both sectors.

\subsection{Even-dilaton couplings}\label{sec:evenD}

Let us now discuss more closely the terms involving an even number of dilatons. Initially, it should be highlighted that for both $\phi^{2}h^{3}$ and $\phi^{4}h$ interactions, we computed the amplitude at tree level and one loop as well as performed the pole subtraction for each case separately. The final result yields exactly the same kinematical structure with identical normalisation for the tree-level and one-loop effective action. While this agreement is fully expected based on general consistency arguments explained in Sec.~\ref{Sec:Method}, it serves as a non-trivial internal check of our computations. This result complements other consistency tests, such as comparisons with the T-duality results of \cite{Garousi:2020gio,Garousi:2020lof} mentioned above and with lower-dimensional supersymmetry \cite{Bonetti:2016dqh} that we discuss below.

There are some subtleties regarding the structure of the effective action due to the choice of field basis. For example, above we argued that the five-point effective action for $\phi^2h^3$ interactions is captured at tree level and one loop by
\begin{equation}\label{eq:TwoDilatonFivePts}
    \cL_{\phi^{2} h^{3}}=\bigl (t_{8}t_{8}-\tfrac{1}{4}\epsilon_{8}\epsilon_{8}\bigr )\, 4R^{3}\bigl [\tfrac{3}{4}\delta\p \phi\p\phi-\tfrac{1}{24}\delta\delta(\p \phi)^{2}\bigr ]\, .
\end{equation}
However, it turns out that there is a two-parameter family of actions because the four-index contractions are not all independent. That is, we find the following identities at the level of the on-shell five-point amplitude
\begin{align}
    \label{eq:IDtwodil}
    \epsilon_{8}\epsilon_{8} R^3\delta\delta(\p \phi)^{2} &= 8 t_{8}t_{8}R^3\delta\delta(\p \phi)^{2}\, , \nn\\[0.4em]
    \epsilon_{8}\epsilon_{8}  R^3\delta\p \phi\p\phi&=-12t_{8}t_{8}R^3\delta\p \phi\p\phi +2 t_{8}t_{8}R^3\delta\delta(\p \phi)^{2}\, .
\end{align}
This implies that \eqref{eq:TwoDilatonFivePts} can equivalently be written solely in terms of $t_8t_8$ as
\begin{equation}
    \cL_{\phi^{2} h^{3}}= t_{8}t_{8}\, R^{3}\bigl [12\delta(\p \phi)^{2}-\tfrac{4}{3}\delta\delta(\p \phi)^{2}\bigl ]
\end{equation}
or in terms of $\epsilon_{8}\epsilon_{8}$ as
\begin{equation}
    \cL_{\phi^{2} h^{3}}=\epsilon_{8}\epsilon_{8}\,  R^{3}\bigl [-\delta(\p \phi)^{2}+\tfrac{1}{12}\delta\delta(\p \phi)^{2}\bigl ]\, .
\end{equation}
These results highlight an obvious, but important point: The structure of the effective action is not uniquely fixed by the amplitude alone, but requires in addition the choice of field basis. In particular, due to identities among index contractions \eqref{eq:IDtwodil} valid on-shell, different but equivalent forms of the five-point $\phi^2 h^3$ couplings can be written using either $t_8 t_8$, $\epsilon_8 \epsilon_8$, or combinations thereof. This illustrates the significance of field redefinitions in determining the form of the effective action and motivates care when comparing effective actions derived from different approaches.

Higher-derivative terms in the 10D effective action can be tested by performing dimensional reduction and comparing the resulting couplings with those allowed by lower-dimensional supersymmetry.
Calabi-Yau threefold reductions, as studied e.g.~in \cite{Antoniadis:1997eg,Becker:2002nn,Bonetti:2016dqh}, provide important guidance for the structure of dilaton couplings in the effective action. Particularly, \cite{Bonetti:2016dqh} proposed the structure of higher-derivative terms of the form $(\nabla \phi)^{2}R^{3}$ by replacing
\begin{equation}\label{eq:BWtest} 
\left (t_{8}t_{8}-\tfrac{1}{4}\epsilon_{8}\epsilon_{8}\right ) R^{4}\rightarrow \left (t_{8}t_{8}-\tfrac{1}{4}\epsilon_{8}\epsilon_{8}\right ) (R^{4}+4\tilde{c}_{2}\delta (\nabla\phi)^{2}R^{3}+4\tilde{c}_{3}\delta \delta (\nabla\phi)^{2}R^{3})
\end{equation}
building on earlier work \cite{Gross:1986mw,Kehagias:1997cq,Kehagias:1997jg}.
Upon restricting \eqref{eq:BWtest} to two-dilaton interactions and reducing the full action on Calabi-Yau threefolds, consistency with $\mathcal{N}=2$ supersymmetry in four dimensions required
\begin{equation}\label{eq:BWconst} 
3\tilde{c}_{2}+6\tilde{c}_{3}-2=0\, .
\end{equation}
By explicitly comparing our five-point results \eqref{eq:TwoDilatonFivePts} with \eqref{eq:BWtest}, we have $\tilde{c}_{2}=3/4$, $\tilde{c}_{3}=-1/24$ and thus the constraint \eqref{eq:BWconst} is satisfied. This provides a non-trivial check on the consistency of our construction and serves as an important cross-validation of the dilaton-curvature couplings derived from the string amplitudes. In this context, let us also note that $\mathcal{N}=2$ supersymmetry in four dimensions can be used to provide another independent argument for the absence in the 10-dimensional effective theory of couplings of the schematic form $f_2 \mathcal{P}^2 R^3$, whose presence would have implied different kinematics at tree level and one loop in Table~\ref{tab:FivePointContact}.

As for the $\phi^4h$ interactions, again there are a large set of possible couplings that are degenerate at the level of the five-point function.  Although it is possible to maintain the standard kinematical structure of this interaction by writing
\begin{equation}
    \mathcal L_{\phi^4h}=\bigl(t_8t_8-\ft14\epsilon_8\epsilon_8\bigr)R(\nabla\nabla\phi)^2\bigl[-\ft1{12}\delta\partial\phi\partial\phi+\ft{61}{1848}\delta\delta(\partial\phi)^2\bigr]\,,
\end{equation}
this form of the coupling, however, does not admit a natural SL(2,$\mathbb{Z}$) invariant generalisation once the RR axion is included.  For this reason, we have chosen a minimal basis for the $\phi^4h$ coupling, as indicated in Table~\ref{tab:FivePointContact}.  This is similar to what happens at the quartic level, where the $\phi^4$ coupling is written directly in terms of the $\mathcal O_1$ operator, (\ref{eq:O1}), and not in terms of contractions with $t_8t_8$ and $\epsilon_8\epsilon_8$.

\subsection{Odd-dilaton couplings}\label{sec:oddD}

The terms with odd powers of the dilaton, specifically $\phi^3 h^2$ and $\phi^5$, exhibit notable structural features. Crucially, we have found a field redefinition basis where the kinematical structure at both tree level and one loop is completely governed by $\epsilon_8 \epsilon_8$, with the only distinction being a relative prefactor.

The relative coefficient provides an important hint towards $\mathrm{SL}(2,\mathbb{Z})$ completion. As mentioned in Sec.~\ref{sec:review}, the corresponding modular-completed couplings maximally violate the $\text{U}(1)$ R-symmetry with the associated modular functions $f_{\pm 1}$ given by \eqref{eq:fpm1}.
From the 4th and 6th line in Table~\ref{tab:FivePointContact}, we notice that the tree and loop coefficients indeed satisfy
\begin{equation}
    \dfrac{\text{tree coeff.}}{\text{loop coeff.}}=-3\,,
\end{equation}
which is as expected from \eqref{eq:fpm1}. Therefore, the contact terms in the odd-dilaton sector can be written in terms of $f_{\pm 1}$ in \eqref{eq:fpm1} as
\begin{equation}
    \cL_{\text{odd}} =\left (f_{1}+f_{-1}\right )\epsilon_{8}\epsilon_{8}\left [ R^{2} (\p\phi)^{2}(\nabla\nabla \phi)+\dfrac{2}{75} (\p \phi)^{2}(\nabla\nabla \phi)^{3}\right ]\,.
\end{equation}
To write down the full $\mathrm{SL}(2,\mathbb{Z})$-invariant form of these couplings, we require additional input from RR-sector couplings which will be the focus of the subsequent section.

\section{RR-sector completion and the effective action}\label{sec:RRcomp}

Having completed the analysis of five-point NSNS graviton and dilaton couplings, we now wish to include the RR axion.  Here, we must proceed with particular care when expressing the effective action in terms of complex fields that include the RR sector. Knowledge of the full NSNS-sector couplings alone is generally insufficient to reconstruct the mixed NSNS/RR interactions. This limitation has been highlighted in earlier analyses of the four-point function, notably in \cite{Policastro:2006vt,Policastro:2008hg}, where it was shown, for example, that
\begin{equation}
t_{8}t_{8}(\nabla\nabla \phi)^{4}\nrightarrow t_{8}t_{8} (|\nabla \cP|^{2})^{2}\, .
\end{equation}
To account for contributions that vanish in the pure NSNS sector, the authors of \cite{Policastro:2008hg} introduced an additional kinematic tensor, $\mathcal{O}_1\big((|\nabla \mathcal{P}|^2)^2\big)$, defined in Eq.~\eqref{eq:O1}. When restricted to the NSNS sector, this tensor reduces to $t_8 t_8 (\nabla\nabla \phi)^4$ at the level of the four-point function.  However, in the mixed sector, it does \textit{not} yield $6t_8t_8(\nabla\nabla\phi)^2(\nabla\nabla\chi)^2$. As we will see shortly, similar to the findings in \cite{Policastro:2008hg}, five-point couplings involving more than two scalars exhibit non-trivial kinematical structures that vanish upon projection to the NSNS subsector.

This behaviour is closely tied to the structure of the full $\mathrm{SL}(2,\mathbb{Z})$-invariant action, which must be built from suitable modular forms. As discussed previously, each term with $n$ scalar insertions is multiplied by a single modular function, independently of the origin of the external states. Consequently, even though new kinematical structures may arise from mixed-sector amplitudes involving RR fields, the total kinematics must factorise consistently at both tree level and one loop.
This implies that it is sufficient to compute a subset of tree-level amplitudes with RR external states to determine the mixed-sector contributions. This is what we will do in Sec.~\ref{sec:RRcontact}. Then, by combining this input with the NSNS-sector results from the previous section, and by using the constraints from S-duality, we are able to reconstruct the complete $\mathrm{SL}(2,\mathbb{Z})$-invariant five-point effective action, which we summarise in its full glory in Sec.~\ref{sec:FinalEFT}.

\subsection{The RR-sector contact terms}\label{sec:RRcontact}

We consider the tree-level two-RR and three-NSNS amplitude.  This closed-string amplitude is obtained by performing the KLT procedure with a two-fermion and three-vector open-string amplitude.  Although this amplitude contains information of all RR fields, we specialise to two axions by taking scalar polarisations for the RR vertices.  The two-axion amplitudes with gravitons and dilatons are then
\begin{equation}
	\chi^2h^3,\qquad \chi^2\phi h^2,\qquad\chi^2\phi^2 h,\qquad\chi^2\phi^3.
\end{equation}
For all of these amplitudes, we isolate the genuine five-point contact terms by the procedure described in Sec.~\ref{Sec:Method}. We will discuss them in turn. The final result is summarised in Table~\ref{tab:FivePoint2chi}.

\begin{table}
\centering
\begin{tabular}{|c||c|}
\hline 
&\\[-0.8em]
 & Tree\\ [0.4em]
 \hline
\hline 
&  \\[-0.8em]
$\chi^{2} h^{3}$ & $\bigl (t_{8}t_{8}-\frac{1}{4}\epsilon_{8}\epsilon_{8}\bigl )\, 4R^{3}\bigl [\frac{3}{4}\delta^{\nu}_{\mu} (\p_{\sigma}\chi)(\p^{\lambda}\chi)-\frac{1}{24}\delta^{\nu}_{\mu}\delta^{\lambda}_{\sigma} (\p \chi)^{2}\bigl ]$   \\ [0.4em]
\hline 
&\\[-0.8em]
$\chi^{2}\phi h^{2}$ & $-2\phi \epsilon_{8}\epsilon_{8}\,  R^{2}(\nabla\nabla \chi)^{2}+\fft3{20}\epsilon_8\epsilon_8R^2((\delta\delta\partial\chi\cdot\partial\chi)(\nabla\nabla\phi)-(\delta\delta\partial\chi\cdot\partial\phi)(\nabla\nabla\chi))$  \\ [0.4em]
\hline 
& \\[-0.8em]
$\chi^{2}\phi^2 h$ &see Eq.~(\ref{eq:c2p2h}) excluding first line\\ [0.4em]
\hline 
&\\[-0.8em]
$\chi^{2}\phi^3$ & see last two lines of Eq.~(\ref{eq:c2p3})  \\ [0.4em]
\hline 
\end{tabular} 
\caption{Genuine five-point contact terms with two axions}\label{tab:FivePoint2chi} 
\end{table}

For the $\chi^2h^3$ amplitude, after subtracting the underlying four-point amplitude, the remaining five-point contact term matches that of $\phi^2h^3$ in Table~\ref{tab:FivePointContact} with dilatons replaced by axions.  Just as in the two-dilaton case, we wish to remove the non-linear component of the quartic effective action.  The result is simply that of $\phi^2h^3$ in Table~\ref{tab:FivePointFunction} with the replacement $\phi\to\chi$.  At the level of two scalars, the amplitude, must be $\mathrm{U}(1)$ preserving.  Hence there is a unique SL(2;$\mathbb{Z}$) completion of the two-dilaton amplitude, with $\partial\phi\partial\phi\to\mathcal P\bar{\mathcal P}$.  Thus there are no unusual kinematics in this case and the two-scalar kinematics is encoded in the ${\rm SL}(2,\mathbb{Z})$-invariant terms
\begin{equation}\label{eq:A2scalar}
    \cL_{2\,\mathrm{scalars}}=(t_8t_8-\tfrac{1}{4}\epsilon_8\epsilon_8)R^3(12\delta \cP\overline{\cP}-\tfrac{2}{3}\delta\delta |\cP|^2)\, .
\end{equation}

For the $\chi^2\phi h^2$ amplitude, the subtraction of the underlying four-point amplitude is actually finite because of the nature of the $\chi\chi\phi$ vertex. After subtraction, we find that the the five-point contact term takes the form
\begin{align}
    \cL^{\mathrm{tree}}_{\chi^2\phi h^2}&=6(t_8t_8-\epsilon_8\epsilon_8)[\fft\phi2(\nabla\nabla\chi)^2R^2+2(\delta\partial\phi\partial\chi)(\nabla\nabla\chi)R^2]\nn\\
    &\quad-2\phi\epsilon_8\epsilon_8R^2(\nabla\nabla\chi)^2+\fft3{20}\epsilon_8\epsilon_8R^2((\delta\delta\partial\chi\cdot\partial\chi)(\nabla\nabla\phi)-(\delta\delta\partial\chi\cdot\partial\phi)(\nabla\nabla\chi)).
\label{eq:ac2ph2}
\end{align}
Note that the first line provides the non-linear completion of the four-point term in (\ref{eq:PTaction}).  To see this, we can explicitly write
\begin{equation}
    2\mathcal P_\mu=\partial_\mu\phi+ie^\phi\partial_\mu\chi\,.
\end{equation}
As a result
\begin{equation}
    4e^{-3\phi/2}R^2\nabla\mathcal P\nabla\bar {\mathcal P}=e^{-3\phi/2}R^2(\nabla\nabla\phi)^2+e^{\phi/2}R^2((\nabla\nabla\chi)^2+2(\delta\partial\phi\partial\chi)(\nabla\nabla\chi)+(\delta\partial\phi\partial\chi)^2)\,.
\label{eq:nldpdp}
\end{equation}
Taking the two-axion part of this, and focusing only on the five-point couplings, we end up with the first line of (\ref{eq:ac2ph2}).  After removing the first line, the second line is then the true five-point contact term.  The first term in the second line is the S-dual version of the $\phi^3h^2$ contact term, while the second term does not have a pure NSNS counterpart.  This is similar to the case of the $\mathcal O_1$ operator in (\ref{eq:PTaction}) that exhibits non-standard kinematics.  However, here it already shows up in the three-scalar amplitude.  Note that, up to a constant factor, this can be written as
\begin{equation}
	\epsilon_8\epsilon_8R^2((\delta\delta\mathcal P_\alpha\bar{\mathcal P}^\alpha)(\nabla\mathcal P)-(\delta\delta\mathcal P_\alpha\mathcal P^\alpha)(\nabla\bar{\mathcal P}))+h.c.\,.
\end{equation}
To summarise, the three-scalar kinematics is captured by the following ${\rm SL}(2,\mathbb{Z})$-covariant couplings
\begin{align}\label{eq:A3scalar}
    \cL_{3\,\mathrm{scalars}} &=\epsilon_8\epsilon_8 R^2\bigl [8(\nabla \cP)(\delta \cP\, \overline{\cP})+\tfrac{3}{10}\bigl((\nabla \cP)(\delta\delta \cP\cdot\overline{\cP})-(\nabla\overline{\cP})(\delta\delta \cP\cdot \cP)\bigr)\bigl ]\, .
\end{align}

For the $\chi^2\phi^2h$ amplitude, in order to subtract the underlying four-point amplitude, we need the form of the $\mathcal O_1$ operator, written in Eq.~\eqref{eq:O1}, in the mixed NSNS/RR sector.  This operator expands at linear order in the scalars as
\begin{align}
    \mathcal O_1\left((|\nabla\cP|^{2})^{2}\right)&=96\bigl[((\nabla_{\mu}\nabla_{\nu} \phi)^2)^2+2\left(2(\nabla_{\mu}\nabla_{\nu}\phi\nabla^{\mu}\nabla^{\nu}\chi)^2-(\nabla_{\mu}\nabla_{\nu}\phi)^2(\nabla_{\sigma}\nabla_{\rho}\chi)^2\right)\nn\\
    &\kern2.5em+((\nabla_{\mu}\nabla_{\nu}\chi)^2)^2\bigr].
\label{eq:O1op}
\end{align}
This expression allows us to subtract the $\phi^2\chi^2$ four-point factorised interaction from the $\chi^2\phi^2h$ amplitude. After doing so, we obtain
\begin{align}
    \cL^{\mathrm{tree}}_{\chi^2\phi^2h}&=96\left[4(\nabla_{\mu}\nabla_{\nu}\phi\nabla^{\mu}\nabla^{\nu}\chi)^2-2(\nabla_{\mu}\nabla_{\nu}\phi)^2(\nabla_{\sigma}\nabla_{\rho}\chi)^2\right]_{5\,\mathrm{pt}}\nn\\[0.2em]
    &\quad+24R^{\mu\nu\rho\sigma}\bigl[-4(\nabla_{\mu}\nabla^{\alpha}\phi)(\nabla_{\sigma}\nabla_{\alpha}\phi)(\nabla_{\nu}\chi)(\nabla_{\rho}\chi)-4(\nabla_{\mu}\nabla^{\alpha}\chi)(\nabla_{\sigma}\nabla_{\alpha}\chi)(\nabla_{\nu}\phi)(\nabla_{\rho}\phi)\nn\\[0.2em]
    &\kern5em+8(\nabla_{\mu}\nabla^{\alpha}\phi)(\nabla_{\sigma}\nabla_{\alpha}\chi)(\nabla_{\nu}\phi)(\nabla_{\rho}\chi)-8(\nabla_{\mu}\nabla^{\alpha}\phi)(\nabla_{\sigma}\nabla_{\alpha}\chi)(\nabla_{\nu}\chi)(\nabla_{\rho}\phi)\nn\\[0.2em]
    &\kern5em+24(\nabla_{\mu}\nabla^{\alpha}\phi)(\nabla_{\nu}\nabla_{\alpha}\chi)(\partial_{\sigma}\phi)(\partial_{\rho}\chi) -24(\nabla_{\mu}\nabla^{\alpha}\phi)(\nabla_{\nu}\nabla_{\alpha}\chi)(\partial_{\sigma}\chi)(\partial_{\rho}\phi)\nn\\[0.2em]
    &\kern5em-19(\nabla_{\mu}\nabla_{\sigma}\phi)(\nabla_{\nu}\nabla_{\rho}\phi)\, (\partial\chi)^2-19(\nabla_{\mu}\nabla_{\sigma}\chi)(\nabla_{\nu}\nabla_{\rho}\chi)\, (\partial\phi)^2\nn\\[0.2em]
    &\kern5em+40(\nabla_{\mu}\nabla_{\sigma}\phi)(\nabla_{\nu}\nabla_{\rho}\chi)(\partial^{\alpha}\phi)(\partial^{\alpha}\chi)\bigr]\,.
\label{eq:c2p2h}
\end{align}
Here, the first line is the five-point contact term obtained by inserting a graviton fluctuation either in an inverse metric contracting the derivatives or in a Christoffel connection contained in the covariant derivative.  This expression is more complicated than its pure NSNS counterpart.  Nevertheless, the four-scalar one-graviton coupling can be written in the ${\rm SL}(2,\mathbb{Z})$-invariant form
\begin{align}\label{eq:A4scalar}
    \cL_{4\,\mathrm{scalars}}&=384R^{\mu\nu\sigma\rho}\bigl(5(\nabla_{\mu}\overline{\cP}_{\sigma})(\nabla_{\nu}\overline{\cP}_{\rho}) \, \cP^{2}+5(\nabla_{\mu}{\cP}_{\sigma})(\nabla_{\nu}{\cP}_{\rho})\, \overline{\cP}\,^{2}-9(\nabla_{\mu}\cP_{\sigma})(\nabla_{\nu}\overline{\cP}_{\rho})\,|\cP|^2\nn\\[0.3em]
    &\hphantom{=384R^{\mu\nu\rho\sigma}\bigl(} -4(\nabla_{\mu}\cP^{\alpha})\overline{\cP}_{\rho}\bigl \{ (\nabla_{\sigma}\overline{\cP}_{\alpha})\, \cP_{\nu}+3(\nabla_{\nu}\overline{\cP}_{\alpha})\, \cP_{\sigma}\bigr \}\bigr)\, .
\end{align}

Finally, the $\chi^2\phi^3$ contact term takes the form
\begin{align}
    \cL_{\chi^2\phi^3}^{\mathrm{tree}}&=96\Bigl[\fft\phi2(4(\nabla_{\mu}\nabla_{\nu}\phi\nabla^{\mu}\nabla^{\nu}\chi)^2-2(\nabla_{\mu}\nabla_{\nu}\phi)^2(\nabla_{\sigma}\nabla_{\rho}\chi)^2)\nn\\
    &\qquad+8(\nabla_{\mu}\nabla_{\nu}\phi\nabla^{\mu}\nabla^{\nu}\chi)(\nabla_{\sigma}\nabla_{\rho}\phi\,\partial^{\sigma}\phi\partial^{\rho}\chi)-4(\nabla_{\mu}\nabla_{\nu}\phi)^2(\nabla_{\sigma}\nabla_{\rho}\chi\,\partial^{\sigma}\phi\partial^{\rho}\chi)\Bigr]\nn\\
    &\qquad+\fft8{75}\phi\epsilon_8\epsilon_8(\nabla\nabla\phi)^2(\nabla\nabla\chi)+\fft4{175}\Bigl(3\epsilon_8\epsilon_8\nabla\nabla\phi(\nabla\nabla\chi)^2(\delta\delta(\partial\phi)^2)\nn\\
    &\kern4em-4\epsilon_8\epsilon_8(\nabla\nabla\phi)^2\nabla\nabla\chi(\delta\delta(\partial\phi)(\partial\chi))+\epsilon_8\epsilon_8(\nabla\nabla\phi)^3(\delta\delta(\partial\chi)^2)\Bigr)\, .
\label{eq:c2p3}
\end{align}
The first two lines correspond to the non-linear completion of $\mathcal O_1$, while the first term in the third line is the counterpart of the five-dilaton contact term.  The five-scalar genuine contact term has a natural ${\rm SL}(2,\mathbb{Z})$-covariant form
\begin{align}\label{eq:A5scalar}
    \cL_{5\,\mathrm{scalars}}&=\epsilon_8\epsilon_8 (\nabla \cP)\bigl [-\tfrac{64}{75}(\nabla \cP)^2(\delta\overline{\cP}\,^2)+\tfrac{16}{175}\bigl\{-3(\nabla \cP)^2(\delta\delta\overline{\cP}\,^2)+2|\nabla \cP|^2(\delta\delta |\cP|^2)\nn\\[0.3em]
    &\hphantom{=\epsilon_8\epsilon_8 (\nabla \cP)\bigl [}+(\nabla\overline{\cP})^2(\delta\delta \cP^2)\bigr\}\bigl ]\,.
\end{align}
In the full effective action, we also have to add the complex conjugate of $\cL_{5\,\mathrm{scalars}}$ multiplied with the appropriate modular function, see Eq.~\eqref{FinalEFT}.

\subsection{The effective action: Final form}\label{sec:FinalEFT}

We are now finally able to write down the effective Type IIB Lagrangian, valid up to five points, for the gravity plus scalar sector. We just need to use the S-duality property of the theory to specify the string-coupling dependence of the various terms. The final expression reads:
\begin{equation}\label{eq:newL}
    \cL=R-2|\cP|^2+\cL^{(4)}+\cL^{(5)},
\end{equation}
where
\begin{align}
    \cL^{(4)}&=\alpha f_0(\tau,\bar\tau)\Bigl[(t_8t_8-\tfrac{1}{4}\epsilon_8\epsilon_8)\bigl (R^4+24R^2|\nabla \cP|^2\bigl )+1536(\nabla_{\mu} \cP_{\nu})^2(\nabla_{\sigma}\overline{\cP}_{\rho})^2\Bigr]\,,
\end{align}
and
\begin{equation}\label{FinalEFT}
    \cL^{(5)} =\alpha f_0(\tau,\bar\tau)\bigl [\cL_{2\,\mathrm{scalars}}+\cL_{4\,\mathrm{scalars}}\bigl ]+\alpha f_1(\tau,\bar\tau)\bigl [\cL_{3\,\mathrm{scalars}}+\cL_{5\,\mathrm{scalars}}\bigl ]+\text{h.c.}\, .
\end{equation}
Here, we collect the $\mathrm{U}(1)$-neutral in
\begin{align}
    \cL_{2\,\mathrm{scalars}} &=(t_8t_8-\tfrac{1}{4}\epsilon_8\epsilon_8)R^3(12\delta \cP\overline{\cP}-\tfrac{2}{3}\delta\delta \cP\cdot\overline{\cP})\, ,\\[0.5em]
    \cL_{4\,\mathrm{scalars}} & = 384R^{\mu\nu\sigma\rho} \bigl( 5(\nabla_{\mu}\overline{\cP}_{\sigma})(\nabla_{\nu}\overline{\cP}_{\rho}) \, \cP^{2}+5(\nabla_{\mu}{\cP}_{\sigma})(\nabla_{\nu}{\cP}_{\rho})\, \overline{\cP}\,^{2}-9(\nabla_{\mu}\cP_{\sigma})(\nabla_{\nu}\overline{\cP}_{\rho})\,|\cP|^2\nn\\[0.3em]
    &\hphantom{=384R^{\mu\nu\rho\sigma}\bigl(} -4(\nabla_{\mu}\cP^{\alpha})\overline{\cP}_{\rho}\bigl \{ (\nabla_{\sigma}\overline{\cP}_{\alpha})\, \cP_{\nu}+3(\nabla_{\nu}\overline{\cP}_{\alpha})\, \cP_{\sigma}\bigr \}\bigr)\, ,
\end{align}
and the maximally $\mathrm{U}(1)$-violating terms in
\begin{align}
    \cL_{3\,\mathrm{scalars}} &=\epsilon_8\epsilon_8 R^2\bigl [8(\nabla \cP)(\delta \cP\, \overline{\cP})+\tfrac{3}{10}\bigl((\nabla \cP)(\delta\delta \cP\cdot\overline{\cP})-(\nabla\overline{\cP})(\delta\delta \cP\cdot \cP)\bigr)\bigl ]\, ,\\[0.5em]
    \cL_{5\,\mathrm{scalars}} &=\epsilon_8\epsilon_8 (\nabla \cP)\bigl [-\tfrac{64}{75}(\nabla \cP)^2(\delta\overline{\cP}\,^2)+\tfrac{16}{175}\bigl\{-3(\nabla \cP)^2(\delta\delta\overline{\cP}\,^2)\nn\\[0.5em]
    &\hphantom{=\epsilon_8\epsilon_8 (\nabla \cP)\bigl [}+2|\nabla \cP|^2(\delta\delta |\cP|^2)+(\nabla\overline{\cP})^2(\delta\delta \cP^2)\bigr\}\bigl ]\, .
\end{align}
This final result provides a concrete realisation of the $\mathrm{SL}(2,\mathbb{Z})$-invariant five-point effective action in the scalar-gravity sector of Type IIB string theory. To recapitulate, by explicitly computing and matching tree-level and one-loop amplitudes involving dilatons, axions, and gravitons, we derived all relevant contact interactions and organised them into modular-covariant structures. The couplings respect the expected $\mathrm{U}(1)$ R-symmetry selection rules, with even-scalar terms governed by the Eisenstein series $f_0(\tau,\bar{\tau})$ and odd-scalar (maximally $\mathrm{U}(1)$-violating) terms paired with $f_{\pm1}(\tau,\bar{\tau})$.

This result not only confirms the consistency of the modular completion with supersymmetry and duality constraints from e.g.~M-theory, but 
will hopefully help to organise the stringy corrections in a more systematic fashion. Given that the main conceptual significance of the F-theory uplift is to provide a geometrisation of the Type IIB axio-dilaton dynamics, it is highly desirable to understand  to which extent such a geometrisation persists beyond leading order in $\alpha'$. Given the appearance of $\mathrm{U}(1)$-violating amplitudes,  the knowledge of higher-derivative corrections beyond four-point interactions is necessary to obtain a complete $\mathrm{SL}(2,\mathbb{Z})$-invariant set of couplings. Incorporation of additional fields such as $F_{5}$ or $G_{3}$ is still an open problem.

The logic of combining somewhat unwieldy stringy interactions to more compact expressions also applies to the M-theory uplift of Type IIA couplings involving graviton, dilaton, and the RR one-form. 
Here, however, due to the existence of a 11D supergravity, it is expected that Type IIA $\alpha'$ corrections are correctly captured by the M-theory higher-derivative dynamics, as we will now describe at the eight-derivative level.

\section{Dilaton couplings and M-theory uplift}\label{sec:IIAMth}

In the above, we have made extensive use of four-point and five-point string amplitudes in creating the effective five-point couplings in the effective Type IIB Lagrangian.  While the tree-level string amplitudes are well established, translating the amplitudes to effective contact terms requires some care, as explained in \S\ref{sec:amptoact}.  In this regard, one important issue that arises is the treatment of the massless external NSNS states, namely ($h_{\mu\nu},b_{\mu\nu},\phi$).  From the string point of view, these closed-string states correspond to a single closed-string vertex operator with different polarisations.  Thus the string amplitude provides a uniform description of graviton, antisymmetric tensor and dilaton scattering.  However, this uniform structure may still yield surprises when translated to the effective Lagrangian.  Since this has led to some confusion when writing down higher-derivative string couplings, we wish to elaborate on the issue below.

In Sec.~\ref{sec:Weyl}, we describe how the dilaton appears in the Type II eight-derivative effective action in the string frame, and in Sec.~\ref{sec:Mtheory} we discuss how this information is encoded in M-theory at eight derivatives.

\subsection{Weyl rescaling to the string frame}\label{sec:Weyl}

To start with, recall that string scattering amplitudes are naturally calculated in the Einstein frame where the graviton and dilaton propagators are diagonal.  Moreover, closed-string vertex operators in the NSNS sector are essentially obtained by combining the left- and right-moving counterparts.  The resulting tensor polarisation has the form
\begin{equation}
    \theta_{\mu\nu}=h_{\nu\nu}+b_{\mu\nu}+\fft12(\eta_{\mu\nu}-k_\mu\bar k_\nu-\bar k_\mu k_\nu)\phi,
\label{eq:thetamunu}
\end{equation}
where $\bar k_\mu$ is an auxiliary momentum satisfying $k\cdot\bar k=1$ and $\bar k\cdot\bar k=0$.   In particular, while the dilaton can be thought of as the trace polarisation, the structure here is to ensure transversality, namely  $k^\mu\theta_{\mu\nu}=0$.  At the level of the NSNS four-point function, each tensor polarisation state $\theta_{\mu\nu}$ has to correspond to a gauge-invariant combination
\begin{equation}
    2k^{[\mu}k_{[\rho}\theta^{\nu]}{}_{\sigma]}\quad\to\quad\tilde R^{\mu\nu}{}_{\rho\sigma}\equiv R^{\mu\nu}{}_{\rho\sigma}+\nabla^{[\mu}H^{\nu]}{}_{\rho\sigma}-\delta^{[\mu}{}_{[\rho}\nabla^{\nu]}\nabla_{\sigma]}\phi.
\label{eq:gaugeicomb}
\end{equation}
As a result, we can directly read off the tree-level eight-derivative Einstein-frame coupling
\begin{equation}
    e^{-1}\cL_{\mathrm{E}}=\alpha e^{-3\phi/2}t_8t_8\tilde R^4=\alpha e^{-3\phi/2}t_8t_8(R+\nabla H-\delta\nabla\nabla\phi)^4.
\end{equation}
It has been observed that the Weyl scaling to the string frame
\begin{equation}
    g_{\mu\nu}^{\mathrm{E}}=e^{-\phi/2}g_{\mu\nu}^{\mathrm{str}},
\label{eq:weylse}
\end{equation}
removes the dilaton completely (at least at linearised order), so the string frame action takes the form of a curvature with torsion
\begin{equation}
    e^{-1}\cL_{\mathrm{str}}=\alpha e^{-2\phi}t_8t_8 R(\Omega_+)^4,
\label{eq:sfnsnsa}
\end{equation}
where $R(\Omega_+)$ is given in Eq.~\eqref{eq:RiemmTWTorsion} and the connection $\Omega_+$ is defined as
\begin{equation}
    \Omega_+^{\alpha\beta}=\omega^{\alpha\beta}+\fft12\mathcal H^{\alpha\beta}\, .
\end{equation}
Here $\omega^{\alpha\beta}=\omega_\mu{}^{\alpha\beta}dx^\mu$ is the torsion-free spin connection, and $\mathcal H^{\alpha\beta}=H_\mu{}^{\alpha\beta}{}dx^\mu$.

That the gauge invariant combination of the Riemann tensor and dilaton, (\ref{eq:gaugeicomb}), precisely matches the Weyl scaling factor between the Einstein and string frames has been used to argue that the dilaton is completely absent from the string-frame effective action.  This is certainly true at the linearised (\textit{i.e.}\ four-point function) order in the NSNS sector.  Furthermore, the absence of the dilaton from the string-frame effective action beyond the linear order is supported by the construction of \cite{Garousi:2020gio,Garousi:2020lof} which uses T-duality to obtain the tree-level $\mathcal O(\alpha'^3)$ effective action.  There it was shown that, in a particular field redefinition frame, the full eight-derivative string-frame action in the NSNS sector can be written completely in terms of the Riemann tensor, $H$, and $\nabla H$, without the need for any dilaton couplings.

Note that demonstrating the absence of dilaton couplings in the string frame is not necessarily straightforward at the non-linear level.  For example, the five-point Einstein-frame graviton and dilaton couplings summarised in Tab.~\ref{tab:FivePointContact} do not have the form of a simple Weyl scaling of the string-frame Riemann tensor.  As a result, converting to the string frame will yield mixed terms with the dilaton.  Nevertheless, as we have demonstrated above, a suitable field redefinition can be used to remove these mixed couplings, resulting in a match with the T-duality results of \cite{Garousi:2020gio,Garousi:2020lof}.

Even at the linear level, the absence of the dilaton in the string-frame effective action is worth a closer look.  Consider, for example, the Einstein-frame Type IIB quartic effective action, \eqref{eq:PTaction}, constructed from the four-point string amplitude in \cite{Policastro:2006vt,Policastro:2008hg}.  Substituting in the individual NSNS and RR fields for $\mathcal P$ and $G_3$ gives
\begin{equation}
    \cL^{(3)}_{\text{4-pt}}=\cL^{(3)}_0+\cL^{(3)}_2+\cL^{(3)}_4,
\end{equation}
where the subscript denotes the number of RR fields.  In the pure NSNS sector, we have
\begin{align}
    \cL^{(3)}_0&=\alpha f_0(\tau,\bar\tau)\left(t_8t_8-\fft14\epsilon_8\epsilon_8\right)\Bigl(R^4+6R^2((\nabla\nabla\phi)^2+(\nabla H)^2)+6(\nabla\nabla\phi)^2(\nabla H)^2\nn\\
    &\kern13em-12R(\nabla\nabla\phi)(\nabla H)^2+((\nabla\nabla\phi)^2)^2+((\nabla H)^2)^2\Bigr).
\label{eq:L03}
\end{align}
This can be rewritten in the compact form
\begin{equation}
    \cL^{(3)}_0=\alpha f_0(\tau,\bar\tau)\left(t_8t_8-\fft14\epsilon_8\epsilon_8\right)\left(R+\nabla H-\nabla\nabla\phi\right)^4,
\end{equation}
where we note that terms with an odd number of $H$ fields vanish by worldsheet parity, and in addition $R^3(\nabla\nabla\phi)$ and $R(\nabla\nabla\phi)^3$ vanish kinematically.  This NSNS expression matches \eqref{eq:gaugeicomb} by construction, indicating that the string-frame action is indeed of the dilaton-free form \eqref{eq:sfnsnsa}.

Turning now to the mixed NSNS and RR couplings, we find
\begin{align}
    \cL_2^{(3)}&=\alpha f_0(\tau,\bar\tau)\biggl[\left(t_8t_8-\fft14\epsilon_8\epsilon_8\right)\Bigl(6(R^2+(\nabla H)^2)(\nabla\nabla\chi)^2+6(R+\nabla\nabla\phi)^2(\nabla F)^2\nn\\
    &\kern13em-24R(\nabla\nabla\chi)\nabla H\nabla F\Bigr)\nn\\
    &\kern5em+192(2(\nabla\nabla\phi\nabla\nabla\chi)^2-(\nabla\nabla\phi)^2(\nabla\nabla\chi)^2)+\mathcal O_2((\nabla H)^2(\nabla F)^2)\biggr].
\end{align}
We can make a couple of observations here.  The first is that the last line corresponds to the non-standard NSNS/RR couplings noted in \cite{Policastro:2008hg}.  After some rearrangement with $t_8t_8$, we can obtain
\begin{align}
    \mathcal L_{\chi^2}&\sim  t_8t_8R^2(\nabla\nabla\chi)^2+32(2(\nabla\nabla\phi\nabla\nabla\chi)^2-(\nabla\nabla\phi)^2(\nabla\nabla\chi)^2)\nn\\
    &=t_8t_8(R-\nabla\nabla\phi)^2(\nabla\nabla\chi)^2-58(\nabla\nabla\phi)^2(\nabla\nabla\chi)^2+12(\nabla\nabla\phi\nabla\nabla\chi)^2\nn\\
    &\quad-24(\nabla\nabla\phi^2\nabla\nabla\chi^2)-12(\nabla\nabla\phi\nabla\nabla\chi\nabla\nabla\phi\nabla\nabla\chi).
\label{eq:twoax}
\end{align}
In particular, the dilaton coupling to the axion is not fully captured by just the first term on the right-hand side, which takes the form $t_8t_8\tilde R^2(\nabla\nabla\chi)^2$.

At first glance, the appearance of such bare dilaton couplings seems to be in tension with the fact that string couplings can be built from the gauge-invariant Riemann combination $\tilde R^{\mu\nu}{}_{\rho\sigma}$ given in (\ref{eq:gaugeicomb}).  The resolution to this, however, is that the couplings are in principle obtained not just from  $\tilde R^{\mu\nu}{}_{\rho\sigma}$, but also from the Ricci combinations
\begin{align}
    \tilde R_{\mu\nu}&=R_{\mu\nu}-\fft12\nabla^\alpha H_{\alpha\mu\nu}-\fft14g_{\mu\nu}\Box\phi-2\nabla_\mu\nabla_\nu\phi,\nn\\
    \tilde R&=R-\fft92\Box\phi.
\label{eq:tildericci}
\end{align}
At linear order, we may impose the equations of motion $R_{\mu\nu}\to0$, $\nabla^\alpha H_{\alpha\mu\nu}\to0$ and $\Box\phi\to0$, in which case we are left with $\tilde R_{\mu\nu}\to-2\nabla_\mu\nabla_\nu\phi$.  What this means is that, at least at linear order, we can arbitrarily trade factors of $\nabla_\mu\nabla_\nu\phi$ with the Ricci combination $\tilde R_{\mu\nu}$.  Another way to see this is to decompose $\tilde R^{\mu\nu}{}_{\rho\sigma}$ into its irreducible components (up to equations of motion)
\begin{align}
    \tilde C^{\mu\nu}{}_{\rho\sigma}&=R^{\mu\nu}{}_{\rho\sigma}+\nabla^{[\mu}H^{\nu]}{}_{\rho\sigma},\nn\\
    \tilde R_{\mu\nu}&=-2\nabla_\mu\nabla_\nu\phi.
\label{eq:Weyltilde}
\end{align}
Here $\tilde C^{\mu\nu}{}_{\rho\sigma}$ is the Weyl tensor constructed from $\tilde R^{\mu\nu}{}_{\rho\sigma}$.  This provides a clean split between the (linearized) torsionful Riemann term that corresponds to $R^{\mu\nu}{}_{\rho\sigma}(\Omega_+)$ in the string frame and the Ricci term that gives rise to $\nabla_\mu\nabla_\nu\phi$ dilaton couplings.  With this in mind, we can express the two-axion couplings, (\ref{eq:twoax}), as
\begin{align}
    \mathcal L_{\chi^2}&=t_8t_8\tilde R^{\mu_1\mu_2}{}_{\nu_1\nu_2}\tilde R^{\mu_3\mu_4}{}_{\nu_3\nu_4}(\delta^{\mu_5}{}_{\nu_5}\nabla^{\mu_6}\nabla_{\nu_6}\chi)(\delta^{\mu_7}{}_{\nu_7}\nabla^{\mu_8}\nabla_{\nu_8}\chi)-\fft{29}2(\tilde R_{\mu\nu})^2(\nabla_\rho\nabla_\sigma\chi)^2\nn\\
    &\quad+3(\tilde R_{\mu\nu}\nabla^\mu\nabla^\nu\chi)^2-6\tilde R_{\mu\nu}\tilde R^{\nu\rho}\nabla_\rho\nabla_\sigma\chi\nabla^\sigma\nabla^\mu\chi-3\tilde R_{\mu\nu}\nabla^\nu\nabla^\rho\chi\tilde R_{\rho\sigma}\nabla^\sigma\nabla^\mu\chi\nn\\
    &=t_8t_8\tilde C^{\mu_1\mu_2}{}_{\nu_1\nu_2}\tilde C^{\mu_3\mu_4}{}_{\nu_3\nu_4}(\delta^{\mu_5}{}_{\nu_5}\nabla^{\mu_6}\nabla_{\nu_6}\chi)(\delta^{\mu_7}{}_{\nu_7}\nabla^{\mu_8}\nabla_{\nu_8}\chi)\nn\\
    &\quad-24\tilde R^{\mu\rho}\tilde C_{\mu\nu\rho\sigma}\nabla^\nu\nabla^\lambda\chi\nabla^\sigma\nabla_\lambda\chi-48\tilde R^{\mu\lambda}\tilde C_{\mu\nu\rho\sigma}\nabla^\rho\nabla_\lambda\chi\nabla^\nu\nabla^\sigma\chi\nn\\
    &\quad-8(\tilde R_{\mu\nu})^2(\nabla_\rho\nabla_\sigma\chi)^2+16(\tilde R_{\mu\nu}\nabla^\mu\nabla^\nu\chi)^2.
\end{align}
Here we have explicitly written out the indices to highlight the contrast between the Riemann and Ricci terms.  At linear order, the mixed $\tilde R\tilde C(\nabla\nabla\chi)^2$ terms cancel against each other, so the quartic two-axion couplings can be written as
\begin{equation}
    \mathcal L_{\chi^2}=t_8t_8\tilde C\tilde C(\nabla\nabla\chi)^2-8(\tilde R_{\mu\nu})^2(\nabla_\rho\nabla_\sigma\chi)^2+16(\tilde R_{\mu\nu}\nabla^\mu\nabla^\nu\chi)^2,
\end{equation}
which is a result that could in fact have been directly obtained by making the substitution (\ref{eq:Weyltilde}) in the first line of (\ref{eq:twoax}).

It should be noted that our choice of $\mathcal O_1\left((|\nabla\mathcal P|^2)^2\right)$, given in (\ref{eq:O1}), differs from that of \cite{Policastro:2006vt,Policastro:2008hg} by a non-trivial field redefinition.  However, regardless of which field redefinition basis is used, the fact remains that Einstein frame couplings to the RR fields will involve the Ricci tensor and Ricci scalar combinations given in (\ref{eq:tildericci}).  These do not show up in the pure NSNS sector, as $t_8t_8$ only contracts indices across different Riemanns.  The Ricci terms in the Einstein frame effective action have the property that they do not contribute to leading order graviton scattering but do contribute to dilaton scattering.  As a result, such terms cannot be determined uniquely by graviton scattering only and can be thought of as effective bare dilaton couplings.

Provided the Einstein frame couplings can be written in terms of $\tilde R^{\mu\nu}{}_{\rho\sigma}$ and its contractions, a Weyl scaling to the string frame will then yield a dilaton-free action, albeit one with Ricci tensor and Ricci scalar terms.  Even in the absence of this Weyl scaling argument, $\nabla\nabla\phi$ terms can always be removed from the string frame effective action by use of the string frame Einstein equation $R_{\mu\nu}=-2\nabla_\mu\nabla_\nu\phi+\cdots$.  Since the dilaton only enters in this form in the leading order action, it can always be removed at this order, regardless of what it couples to.  However, this may no longer be the case at non-linear order where bare $\phi$ and $\partial_\mu\phi$ couplings will start to show up.

The second observation is similar in that the coupling to the RR three-form $F$ takes the form $(R+\nabla\nabla\phi)^2(\nabla F)^2$ with a relative plus sign between Riemann and $\nabla\nabla\phi$.  Since this is the wrong sign compared with (\ref{eq:gaugeicomb}), it yields the string frame coupling
\begin{align}
    \cL_{\mathrm{str}}&=\alpha e^{-2\phi}\left(t_8t_8-\fft14\epsilon_8\epsilon_8\right)\left(6(R+2\nabla\nabla\phi)^2(\nabla F)^2+\cdots\right)\nn\\
    &=\alpha e^{-2\phi}\left(t_8t_8-\fft14\epsilon_8\epsilon_8\right)\bigl(\cdots+24R(\nabla\nabla\phi)(\nabla F)^2+\cdots\bigr).
\end{align}
Again, we could perform a field redefinition to replace $\nabla\nabla\phi$ by the Ricci tensor, leading to a dilaton-free string frame action.  However, it is instructive to consider the field redefinition frame where we keep $\nabla\nabla\phi$ and avoid Riccis.  In this case, while here $F$ is the RR three-form field strength in Type IIB theory, this string-frame coupling is the T-dual version of the one found recently in \cite{Aggarwal:2025lxf} from the circle reduction of M-theory, namely $R(\nabla\nabla\phi)(\nabla F_2)^2$.  We will examine this IIA term in more detail in the next subsection.

Finally, for completeness, we have
\begin{equation}
    \cL_4^{(3)}=\alpha f_0(\tau,\bar\tau)\left(t_8t_8-\fft14\epsilon_8\epsilon_8\right)\left(((\nabla\nabla\chi)^2)^2+6(\nabla\nabla\chi)^2(\nabla F)^2+((\nabla F)^2)^2\right).
\end{equation}
This expression can be obtained by dropping the Riemann tensor from the NSNS Lagrangian, (\ref{eq:L03}), and taking the map $\phi\to\chi$ and $H\to F$.

\subsection{M-theory reduction and Type IIA}\label{sec:Mtheory}

Having seen that the dilaton indeed couples to RR fields in the IIB string-frame action in a Ricci-free field redefinition basis, T-duality indicates that the same will happen in the IIA case.  However, this can also be seen by direct computation.  In the M-theory case, it is expected that the one-loop $t_8t_8\hat{R}^4$ coupling in terms of the eleven-dimensional Riemann tensor $\hat{R}$ will reduce to the corresponding one-loop Type IIA coupling.  Upon performing a standard Kaluza-Klein reduction
\begin{align}
    \mathrm{d}s_{11}^2&=\mathrm{e}^{-\phi/6}\mathrm{d}s_{10}^2+\mathrm{e}^{4\phi/3}(\mathrm{d}y+C_1)^2,\nn\\
    C_3&=C_3+B_2\wedge(\mathrm{d}y+C_1),
\end{align}
the two-derivative bosonic 11-dimensional Lagrangian
\begin{equation}
    e^{-1}\cL_{11}=\hat R-\fft1{2\cdot4!}G_4^2+\fft16C_3\wedge G_4\wedge G_4,
\end{equation}
reduces to the bosonic Type IIA Lagrangian in the Einstein frame
\begin{equation}
    e^{-1}\cL_{10}=R-\fft12\partial_\mu\phi^2-\fft1{12}e^{-\phi}H_{\mu\nu\rho}^2-\fft14e^{3\phi/2}F_{\mu\nu}^2-\fft1{2\cdot4!}e^{\phi/2}F_{\mu\nu\rho\sigma}^2+\fft12B_2\wedge F_4\wedge F_4.
\end{equation}
At the eight-derivative level, it is easy to see that $t_8t_8\hat R^4$ in 11 dimensions reduces to
\begin{equation}
    e^{-1}\cL_{\partial^4}=e^{\phi/2}\left(t_8t_8R^4+\cdots\right),
\end{equation}
where the dilaton factor indicates that this is a one-loop Type IIA term.  The terms in the ellipses denote additional couplings to the dilaton and RR two-form field strength $F_2$.  Additionally, the reduction of 11-dimensional couplings involving $G_4$ will give rise to $H_3$ and $F_4$ couplings in the Type IIA Lagrangian.

The full reduction of 11-dimensional $t_8t_8\hat R^4$ gives rise to a large set of terms, including one-loop $R(\nabla\nabla\phi)(\nabla F_2)^2$, as mentioned above and discussed recently in \cite{Aggarwal:2025lxf}.  While this is a one-loop term, one can make the argument that terms with similar kinematics should exist at tree level as well.  This was shown by direct computation for four-point NSNS tree and loop amplitudes, while for RR fields, one can make the argument that $\mathrm{U}(1)$ preservation at the four-point level ensures that Type IIB kinematics are identical.  Since Type IIA is T-dual to Type IIB this demonstrates that tree and loop equality holds for Type IIA as well.

It is straightforward to verify the existence of a tree-level $R(\nabla\nabla\phi)(\nabla F_2)^2$ coupling (and hence a one-loop coupling) from a direct string four-point amplitude calculation.  Since we want to couple to two $F_2$ fields, the string amplitude is a mixed one with two NSNS and two RR fields.  For the NSNS vertices, we focus only on the graviton and dilaton using the polarisation (\ref{eq:thetamunu}) with $b_{\mu\nu}$ set to zero.  In this way, we can study the scattering of $h^2F_2^2$, $h\phi F_2^2$ and $\phi^2F_2^2$.

In order to compare the string amplitudes with the corresponding M-theory reduction, we make use of the \textit{linearised} Riemann reduction
\begin{align}\label{eq:Rhatred}
    \hat R^{\mu\nu}{}_{\rho\sigma}&=R^{\mu\nu}{}_{\rho\sigma}+\fft13\delta^{[\mu}{}_{[\rho}\nabla^{\nu]}\nabla_{\sigma]}\phi\, ,\nn\\
    \hat R^{10\,\nu}{}_{\rho\sigma}&=\fft12\nabla^\nu  F_{\rho\sigma}\, ,\nn\\
    \hat R^{10\,\nu}{}_{10\,\sigma}&=-\fft23\nabla^\nu\nabla_\sigma\phi\, .
\end{align}
Note that we work here in the Einstein frame as it is more appropriate for direct comparison with the string amplitudes.  The reduction of $t_8 t_8\hat R^4$ yielding two $F_2$ field strengths will have to incorporate two 11-dimensional Riemann tensors of the form $\hat R^{10\,\nu}{}_{\rho\sigma}$. We find
\begin{equation}
    \left.t_8 t_8 \hat R^4\right|_{F^2}=\mathcal T_0+\mathcal T_1+\mathcal T_2,
\end{equation}
where $\mathcal T_{n}$ corresponds to the insertion of $n$ Riemann tensors of the form $\hat R^{10\,\nu}{}_{10\,\sigma}$.  A straightforward calculation yields
\begin{align}
    \mathcal T_0&=12t_8^{\mu_1\cdots\mu_8}\left(\hat R^{\alpha\beta}{}_{\mu_1\mu_2}\hat R^{\beta\alpha}{}_{\mu_3\mu_4}\nabla^\gamma F_{\mu_5\mu_6}\nabla^\gamma F_{\mu_7\mu_8}-4\hat R^{\alpha\beta}{}_{\mu_1\mu_2}\hat R^{\beta\gamma}{}_{\mu_3\mu_4}\nabla^\alpha F_{\mu_5\mu_6}\nabla^\gamma F_{\mu_7\mu_8}\right),\nn\\
    \mathcal T_1&=-64\Bigl(\hat R^{\mu\nu\gamma\delta}\nabla^\alpha\nabla^\beta\phi\nabla^\alpha F_{\mu\nu}\nabla^\beta  F_{\gamma\delta}-4\hat R^{\mu\nu\beta\delta}\nabla^\alpha\nabla_\beta\phi\nabla^\gamma F_{\mu\nu}\nabla^\alpha  F_{\gamma\delta}\nn\\
    &\kern4em+4\hat R_{\alpha\delta\beta\nu}\nabla^\alpha\nabla^\beta\phi\nabla^\gamma F_{\mu\nu}\nabla^\mu F_{\gamma\delta}+4\hat R_{\mu\beta\gamma\alpha}\nabla^\delta\nabla^\nu\phi\nabla^\alpha F_{\mu\nu}\nabla^\beta F_{\gamma\delta}\nn\\
    &\kern4em-8\hat R_{\beta\delta\nu\alpha}\nabla^\mu\nabla^\beta\phi\nabla^\gamma F_{\mu\nu}\nabla^\alpha F_{\gamma\delta}-4\hat R_{\mu\nu\gamma\alpha}\nabla^\delta\nabla^\beta\phi\nabla^\alpha F_{\mu\nu}\nabla^\beta F_{\gamma\delta}\Bigr),\nn\\
    \mathcal T_2&=\fft{128}3\Bigl(-\nabla^\alpha\nabla_\mu\phi\nabla^\alpha\nabla_\mu\phi\nabla^\beta F_{\nu\rho}\nabla^\beta F_{\nu\rho}-2\nabla^\alpha\nabla_\mu\phi\nabla^\beta\nabla_\mu\phi\nabla^\alpha F_{\nu\rho}\nabla^\beta F_{\nu\rho}\nn\\
    &\kern4em+4\nabla^\alpha\nabla_\mu\phi\nabla^\alpha\nabla_\nu\phi\nabla^\beta F_{\mu\lambda}\nabla^\beta F_{\nu\lambda}+4\nabla^\alpha\nabla_\mu\phi\nabla^\beta\nabla_\nu\phi\nabla^\alpha F_{\nu\lambda}\nabla^\beta F_{\mu\lambda}\nn\\
    &\kern4em+4\nabla^\alpha\nabla_\mu\phi\nabla^\beta\nabla_\nu\phi\nabla^\alpha F_{\mu\lambda}\nabla^\beta F_{\nu\lambda}\Bigr).
\end{align}
Note that here $\hat R_{\mu\nu\rho\sigma}$ is the ten-dimensional components of the 11-dimensional Riemann tensor, corresponding to the first line of (\ref{eq:Rhatred}).  Furthermore, because of the nature of $t_8t_8$, there are no Ricci-like contractions of $\hat R_{\mu\nu\rho\sigma}$.

Comparing with the corresponding four-point string amplitudes, we find perfect agreement in the sense
\begin{align}
    \cL_{h^2F^2}&=\left.\mathcal T_0\right|_{\hat R\to R},\nn\\
    \cL_{h\phi F^2}&=\left.\mathcal T_0\right|_{\hat R^2\to2R(\fft13\nabla\nabla\phi)}+\left.\mathcal T_1\right|_{\hat R\to R},\nn\\
    \cL_{\phi^2F^2}&=\left.\mathcal T_0\right|_{\hat R^2\to(\fft13\nabla\nabla\phi)^2}+\left.\mathcal T_1\right|_{\hat R\to\fft13\nabla\nabla\phi}+\mathcal T_2.
\end{align}
Here, the left-hand side expressions are obtained from the corresponding closed-string four-point amplitudes while the right-hand side expressions are those obtained from the Kalua-Klein reduction from 11 dimensions.  Finally, note that the substitution $\hat R\to\lambda\nabla\nabla\phi$ corresponds to a Weyl scaling of the metric for some factor $\lambda$.  For the term in question, $\cL_{h\phi F^2}\sim R\nabla\nabla\phi(\nabla F_2)^2$, the four-point functions satisfy the kinematic identity
\begin{equation}
    \left.\mathcal T_1\right|_{\hat R\to R}=8\left.\mathcal T_0\right|_{\hat R^2\to2R(\fft13\nabla\nabla\phi)}.
\end{equation}
As a result, the string amplitude can also be written as
\begin{equation}
    \cL_{h\phi F^2}=3\left.\mathcal T_0\right|_{\hat R^2\to2R\nabla\nabla\phi}.
\end{equation}
This yields the combination
\begin{align}
    \cL_{h^2F^2}+\cL_{h\phi F^2}+\cL_{\phi^2F^2}=\left.\mathcal T_0\right|_{\hat R\to R+3\nabla\nabla\phi}+\cdots.
\end{align}
What this indicates is that the Einstein-frame coupling to two $F_2$ field strengths can be written using a shifted Riemann tensor
\begin{equation}
    R^{\mu\nu}{}_{\rho\sigma}+3\delta^{[\mu}{}_{[\rho}\nabla^{\nu]}\nabla_{\sigma]}\phi=\tilde R^{\mu\nu}{}_{\rho\sigma}+4\delta^{[\mu}{}_{[\rho}\nabla^{\nu]}\nabla_{\sigma]}\phi\quad\to\quad\tilde R^{\mu\nu}{}_{\rho\sigma}-2\delta^{[\mu}{}_{[\rho}\tilde R^{\nu]}{}_{\sigma]},
\end{equation}
where the final expression here has made use of the leading-order equation of motion for $\tilde R_{\mu\nu}$.  As a result, a transformation to the string frame will give rise to either Ricci or dilaton terms.  In particular, in a Ricci-free field redefinition basis, the $R\nabla\nabla\phi(\nabla F_2)^2$ coupling will survive.

While we have directly shown the existence of a $R\nabla\nabla\phi(\nabla F_2)^2$ coupling in the IIA string effective action, T-duality of the IIB $R(\nabla\nabla\phi)(\nabla F_3)^2$ coupling also yields a IIA coupling of the form $R(\nabla\nabla\phi)(\nabla F_4)^2$.  Although we have not made an explicit string-amplitude check, we have no reason to doubt that such a term would also be present in the string-frame action.  Note that T-duality allows for many possible couplings, although not all are present kinematically.  For example, T-dualising the IIA coupling $R(\nabla\nabla\phi)(\nabla F_2)^2$ suggests the existence of a IIB coupling of the form $R(\nabla\nabla\phi) (\nabla\nabla\chi) (\nabla F_3)$.  However, one can verify that this would vanish kinematically, which is consistent with its absence in the IIB effective action, (\ref{eq:PTaction}).

As demonstrated in \cite{Policastro:2006vt}, the tree-level quartic string effective action can be written in the Einstein frame purely in terms of $\tilde R^{\mu\nu}{}_{\rho\sigma}$ and the RR field strengths $F_{(n)}$.  Once Weyl scaled into the string frame, the resulting string frame action will then be free of explicit dilaton couplings, but will include Ricci tensor and Ricci scalar terms.  This raises the question of whether the dilaton can also be removed from the string frame action beyond quartic order.  The answer is affirmative in the pure NSNS sector, as demonstrated using T-duality invariance in \cite{Garousi:2020gio,Garousi:2020lof}.  (This has also been argued from the string sigma model perspective in \cite{Tseytlin:2006ak}.)  Evidence for the absence of string-frame dilaton couplings at one-loop can also be seen from heterotic/IIA duality \cite{Liu:2013dna}.  However, it is unlikely to be the case with RR couplings beyond linear order, as the RR fields naturally include a factor of $e^\phi$ \cite{Witten:1995ex,Bergshoeff:1995as,Tseytlin:1996hs}.  This can be seen, for example, in the RR axion expression
\begin{equation}
    2\nabla_\mu\mathcal P_\nu=\nabla_\mu(\nabla_\nu\phi+ie^\phi\nabla_\nu\chi)=\nabla_\mu\nabla_\nu\phi+ie^{\phi}(\nabla_\mu\nabla_\nu\chi+\partial_\mu\phi\partial_\nu\chi).
\end{equation}
The non-linear term $\sim\partial\phi\partial\chi$ does not contribute to four-point functions, but must be included at higher orders as we have already seen in (\ref{eq:nldpdp}).

\section{Discussion}\label{sec:Concl}

In this paper we have derived the exact-in-$g_s$ Type IIB eight-derivative effective action up to five points in the gravity plus scalar sector. Prominent features include: Same kinematics for tree level and one loop, $\mathrm{U}(1)$ invariance of couplings with even number of scalars vs maximal $\mathrm{U}(1)$ violation of couplings with odd number of scalars, full agreement with T-duality at tree level in the NSNS sector, appearance of kinematical structures other than $t_8t_8$ and $\epsilon_{10}\epsilon_{10}$ starting from the three-scalar sector.

We have also clarified that there is an interplay between the dilaton and Ricci terms in the string-frame effective action.  Using the string-frame Einstein equation, one can trade $\nabla_\mu\nabla_\nu\phi$ for $R_{\mu\nu}$ or vice versa.  However, one cannot remove both simultaneously, except in the pure NSNS sector.  Taking this into account, we have verified that one-loop $(\alpha')^3$ couplings in Type IIA are correctly captured by the circle reduction of eight-derivative terms in M-theory, at least at quartic order.

There is clearly room for many further investigations. Let us highlight a few possible directions.

\paragraph{Inclusion of $G_3$.} It would be highly desirable to extend the analysis presented in \cite{Liu:2022bfg}, as well as in the present work, to include five-point scattering amplitudes involving all possible interactions among the graviton, the axio-dilaton, and the full $\mathrm{SL}(2,\mathbb{Z})$ doublet of two-form potentials, $B_2$ and $C_2$. Such an extension would allow for a systematic derivation of the complete set of eight-derivative couplings in the bosonic sector of Type IIB string theory. In particular, it would enable a unified treatment of NSNS and RR interactions at higher orders in $\alpha'$, and further clarify how $\mathrm{SL}(2,\mathbb{Z})$ symmetry constrains the effective action beyond the four-point level.

\paragraph{Systematics of pole subtraction.} As we have seen, translating scattering amplitudes into local effective couplings in the Lagrangian is a subtle and often non-trivial process. This challenge becomes increasingly severe as the number of external states grows, due to the rapid increase in factorisation channels and the corresponding complexity of intermediate exchange diagrams. At higher multiplicities, disentangling genuine contact terms from contributions arising from lower-point interactions requires a careful and systematic subtraction of poles, both singular and finite. Without a well-defined procedure, this can obscure the identification of new higher-derivative operators and lead to ambiguities in the effective action. To address this, developing a systematic and generalisable framework for pole subtraction that can be consistently applied to higher-point amplitudes would be an important step forward. Such a framework would not only streamline the derivation of local couplings at six points and beyond, but also clarify how string amplitudes encode the underlying structure of the effective theory. Ultimately, it would provide the essential technical foundation for completing the higher-derivative expansion of string effective actions across a broad range of backgrounds and configurations.

\paragraph{Supersymmetry and higher-dimensional index structures.} While the use of higher-dimensional tensor structures in Type IIB supergravity is well established \cite{Howe:1983sra, deHaro:2002vk, Green:2003an, Liu:2022bfg} --- particularly at the eight-derivative level through the linearised scalar superfield formalism developed in \cite{Howe:1983sra} --- their full role in the non-linear ten-dimensional effective action remains only partially understood. Hereby, a chiral scalar superfield describes the full supergravity multiplet, and serves as a natural object for organising higher-derivative corrections. At the linearised level, this superfield has been used, for example, to construct the supersymmetric completion of the $R^4$ coupling \cite{Green:2003an}. However, much less is known about how such superfield-based structures generalise in the presence of non-trivial backgrounds involving spacetime-dependent axio-dilaton gradients $\mathcal{P}$. Their description requires an extension of the superfield formalism beyond linear order. Exploring this generalisation is essential for understanding the complete structure of the Type IIB effective action from the perspective of supersymmetry.

\paragraph{$\alpha'$ corrections in lower dimensions.} A powerful approach to studying $\alpha'$ corrections in lower-dimensional string compactifications is to perform dimensional reductions of higher-derivative terms in the ten-dimensional effective action. By performing dimensional reductions of these higher-derivative terms on compact spaces such as K3 surfaces or Calabi–Yau threefolds, or even on manifolds with positive Ricci curvature, to account for non-trivial axio-dilaton vacuum profiles, one can in principle systematically extract the corresponding corrections to the various sectors in six or four dimensions. As demonstrated in \cite{Liu:2022bfg}, these reductions translate the $\mathrm{SL}(2,\mathbb{Z})$-invariant structures of the ten-dimensional theory, including the axio-dilaton dependence and modular functions, into couplings involving the moduli fields of the compactification. For example, understanding how the three-scalar interactions contribute upon dimensional reduction remains an open problem. Their role in generating effective couplings in lower dimensions is not yet fully understood and warrants further investigation.

\paragraph{Uplift to F-theory?} Building on the idea that the two-derivative action for 10D Einstein gravity and the axio-dilaton arises from 12D Einstein gravity via an auxiliary torus, \cite{Minasian:2015bxa} showed that 12D covariance also reproduces the eight-derivative four-point effective action in the scalar–graviton sector. It is therefore natural, in our present context, to ask whether this correspondence extends beyond four points. However, the mere presence of $\mathrm{U}(1)$-violating couplings at five points implies that a straightforward 12D uplift as proposed in \cite{Minasian:2015bxa} cannot work beyond four points. Still, one might hope that five-point couplings with an even number of scalars, which instead we have shown to always be $\mathrm{U}(1)$ preserving, may be captured by 12D covariance. But even in this case, we find a mismatch at the level of the five-point function as explained in App.~\ref{app:Ftheory}, see Eq.~\eqref{Mismatch}. On closer inspection, this mismatch might not come unexpected. In the context of deriving 10D couplings from a Schwinger-loop computation in 11D \cite{Green:1997as,Green:1999by}, it becomes clear that the torus reduction used in \cite{Minasian:2015bxa} captures only the contributions from loop amplitudes without insertions of Kaluza-Klein modes of the 11D superparticle. To obtain the full picture, the missing contributions, including those associated with $\mathrm{U}(1)$-violating interactions, must be properly incorporated. Clarifying their origin and exploring the possibility of a twelve-dimensional uplift that captures the complete ten-dimensional effective action remain important open questions. We leave a detailed investigation of this direction for future work.

\section*{Acknowledgements}

We would like to thank Manki Kim for useful discussions and Linus Wulff for correspondence.
AS thanks DAMTP at the University of Cambridge for hospitality where parts of this work have been completed. JTL acknowledges the hospitality of IPhT CEA/Saclay and LPTHE Jussieu. RS thanks the Department of Physics at Cornell University for hospitality at the early stage of this project. This work was supported in part by the US Department of Energy under grant DE-SC0007859 (JTL).

\appendix

\section{From string scattering amplitudes to effective actions}
\label{app:strings}

In perturbative string theory, higher-derivative interactions in the low-energy effective action are most directly extracted by expanding on-shell scattering amplitudes at low momentum.  For example, the tree-level four-graviton amplitude in Type II superstring theory admits a well-defined $\alpha'$ expansion whose first non-trivial correction beyond the two-derivative Einstein term appears at order $(\alpha')^3$. Matching this on-shell expression to a spacetime Lagrangian yields the familiar tree-level $R^4$ interaction, along with its precise index contractions, first derived in \cite{Gross:1986iv,Gross:1986mw}. Supersymmetry and S-duality then extend this result to the full $\mathcal{N}=2$ superinvariant that incorporates the dilaton and three-form corrections at the same order.

In this section, we begin by reviewing key aspects of the NSNS sector, which will serve as our main focus throughout. We discuss the structure of four-, and five-point functions at both tree level and one loop, and summarise how these amplitudes have been used in the literature to reconstruct the corresponding effective action at the eight-derivative level. Building on this, we outline how to extend these results to enable systematic pole subtraction in the scalar-graviton sector, which is essential to isolate genuine\footnote{As stressed in Sec.~\ref{Sec:Method}, one must also remove the non-linear completion of four-point contact terms to extract the truly five-point couplings. However, the latter procedure is much more straightforward than the pole subtraction, and thus we will not discuss its details here.} contact interactions involving dilatons.

While our primary emphasis is on the NSNS sector, many of the techniques and intermediate results also apply to mixed-sector amplitudes involving RR fields. We comment on these connections at the end of the section.

\subsection{Tree level and one-loop amplitudes in the NSNS sector up to five points}

We begin with a review of the tree-level and one-loop amplitudes in the NSNS sector up to five points. That is, we focus on scattering amplitudes with external NSNS states, namely the metric, antisymmetric tensor and dilaton ($h_{\mu\nu},b_{\mu\nu},\phi$).

As demonstrated by Kawai-Lewellen-Tye (KLT) relations \cite{Kawai:1985xq}, closed string amplitudes may be obtained from products of open string amplitudes. Thus, even though we are ultimately interested in closed string amplitudes, the basic building blocks are given by open string amplitudes. More explicitly, the $n$-point closed string amplitude $ \mathcal{M}_n = \mathcal{M}(1,\ldots ,n)$ can be written in terms of $n$-point open string amplitudes $\mathcal{A}(1,\ldots,n)$, $\widetilde{\mathcal{A}}(1,\ldots,n)$ as \cite{Kawai:1985xq}
\begin{equation}\label{eq:KLT}
     \mathcal{M}_n= \sum_{\text{permutations}}\,\widetilde{\mathcal{A}}(1,\ldots,n)\cdot \mathcal{S}_{n}\cdot\mathcal{A}(1,\ldots,n)\, ,
\end{equation}
where $\mathcal{S}_{n}$ is called the KLT kernel which depends generically on the momenta.

By virtue of the KLT relations, the closed string NSNS vertex operator itself becomes a combination of the left- and right-movers where the closed string polarisations can be taken as tensor products of open-string polarisations $e_\mu$,
\begin{equation}\label{eq:NSNSvertex} 
\theta_{\mu\nu}= e_\mu\otimes\tilde{e}_\nu\, .
\end{equation}
The various NSNS fields are simply encoded as
\begin{equation}
\theta^{\mu\nu}=h^{\mu\nu}+b^{\mu\nu}+\ft12(\eta^{\mu\nu}-\bar k^\mu k^\nu-k^\mu\bar k^\nu)\phi.
\label{eq:thetapol}
\end{equation}
Here the metric is expanded as $g_{\mu\nu} = \eta_{\mu\nu} + h_{\mu\nu}$, with $h_{\mu\nu}$ taken to be transverse and traceless, i.e., $k^\mu h_{\mu\nu} = 0$ and $h^\mu{}_\mu = 0$. We note that this specific choice of polarisations in \eqref{eq:thetapol} yields scattering amplitudes in the Einstein frame. The antisymmetric tensor field $B_2 =\tfrac{1}{2} b_{\mu\nu} \mathrm{d} x^{\mu}\wedge\mathrm{d} x^{\nu}$ is described by $b_{\mu\nu}$, with field strength $H_3 = \mathrm{d}B_2$, and satisfying the transversality condition $k^\mu b_{\mu\nu} = 0$. The scalar $\phi$ denotes the canonically normalised dilaton. The auxiliary vector $\bar{k}^\mu$ is defined such that $\bar{k}^2 = 0$ and $k \cdot \bar{k} = 1$, which ensures the transversality condition $k_\mu \theta^{\mu\nu} = 0$ for the dilaton polarisation tensor.

We begin by reviewing the well-known result for the tree-level three-point function. For each massless open string vector, we label its polarisation index by $\mu_i$ and its momentum by $k_i$. The corresponding three-point function in open superstring theory reproduces the standard trilinear gluon amplitude
\begin{equation}
\mathcal{A}(1,2,3)= \left (\eta^{\mu_1\mu_2}k_{12}^{\mu_3}+\eta^{\mu_2\mu_3}k_{23}^{\mu_1}
+\eta^{\mu_3\mu_1}k_{31}^{\mu_2} \right ) e_{1,\mu_{1}}\, e_{2,\mu_{2}}\, e_{3,\mu_{3}} \, ,
\end{equation}
where we have defined $k_{ij} = k_{i}-k_{j}$ and ignored overall factors of couplings and normalisations.
The corresponding expressions $\widetilde{\mathcal{A}}(1,2,3)$ for the right-movers is obtained upon replacing $e_{i,\mu_{i}}\rightarrow \tilde{e}_{i,\mu_{i}}$.
The closed string three-point amplitude $\mathcal{M}(1,2,3)$ is constructed via the KLT relation \eqref{eq:KLT} for which $\mathcal{S}_{3}$ is trivial and therefore 
\begin{equation} \label{eq:M3}
\mathcal{M}_3 = \widetilde{\mathcal{A}}(1,2,3)\mathcal{A}(1,2,3) \, .
\end{equation}
Recalling the definition of the closed string polarisations \eqref{eq:NSNSvertex}, this can be written as
\begin{align}\label{eq:c3pt}
	\mathcal{M}_3 &= 4k_1^{\mu_3}k_1^{\nu_3}\theta_{1, \mu_1\nu_1}\theta_{2,\mu_1\nu_1} \theta_{3,\mu_3\nu_3}-4k_2^{\mu_3}k_2^{\nu_1}\theta_{1,\mu_1\nu_1}\theta_{2,\mu_1\nu_2}\theta_{3,\mu_3\nu_2}\nonumber\\
		&\quad -4k_1^{\mu_3}k_1^{\nu_2}\theta_{1,\mu_1\nu_1}\theta_{2,\mu_1\nu_2} \theta_{3,\mu_3\nu_1}+\mbox{cyclic}\, ,
\end{align}
which is the same for both the IIA and IIB superstring.

Next, let us look at the four-point functions in the NSNS sector. As before, we start with the open string four-point expressions  which have been known at tree level \cite{Green:1981xx} and one-loop \cite{Green:1981ya} for a long time. In fact, they have both the same kinematical structure corresponding for massless external states to the Yang-Mills four-point amplitude
\begin{equation}
\mathcal{A}(1,2,3,4)=-\fft{t_8(k_1,e_1,k_2,e_2,k_3,e_3,k_4,e_4)}{s_{12}\, s_{23}}\, ,
\end{equation}
where $s_{ij}=k_i\cdot k_j$. 
Here we have introduced
\begin{equation} 
t_8(\zeta_1,\zeta_2,\ldots,\zeta_8)=t_{8\,\mu_1\mu_2\cdots\mu_8}\zeta_1^{\mu_1}\zeta_2^{\mu_2}\cdots\zeta_8^{\mu_8},
\end{equation}
in terms of the $t_8$ tensor given by \cite{Schwarz:1982jn}
\begin{align}\label{eq:t8}
t_{8\,\mu_1\nu_1\cdots\mu_4\nu_4}&=-2\bigl((\eta_{\nu_2\mu_1}\eta_{\nu_1\mu_2})(\eta_{\nu_4\mu_3}\eta_{\nu_3\mu_4})+(\eta_{\nu_3\mu_1}\eta_{\nu_1\mu_3})(\eta_{\nu_4\mu_2}\eta_{\nu_2\mu_4})+(\eta_{\nu_4\mu_1}\eta_{\nu_1\mu_4})(\eta_{\nu_3\mu_2}\eta_{\nu_2\mu_3})\bigr)\nn\\
&\quad+8\bigl(\eta_{\nu_4\mu_1}\eta_{\nu_1\mu_2}\eta_{\nu_2\mu_3}\eta_{\nu_3\mu_4}+\eta_{\nu_4\mu_1}\eta_{\nu_1\mu_3}\eta_{\nu_3\mu_2}\eta_{\nu_2\mu_4}+\eta_{\nu_2\mu_1}\eta_{\nu_1\mu_3}\eta_{\nu_3\mu_4}\eta_{\nu_4\mu_2}\bigr )\, .
\end{align}
Note that the right-hand side of this expression is implicitly antisymmetrised in all index pairs $[\mu_i\nu_i]$.
Each external state is massless and on-shell, characterised by momentum $k_i$ and polarisation $e_i$, which obey $k_i^2 = 0$, $k_i \cdot e_i = 0$, and the momentum conservation condition $k_1 + k_2 + k_3 + k_4 = 0$.

As before, we use KLT relations \eqref{eq:KLT} to obtain the tree-level closed string four-point functions.
At tree-level, the closed string four-point function takes the form \cite{Gross:1986iv}
\begin{equation}
\mathcal{M}^{\mathrm{tree}}_{4}\sim\left(\fft8{s_{12}s_{23}s_{13}}+\alpha'^32\zeta(3)+\cdots\right)\, \bigl |s_{12}s_{23}\, \mathcal{A}(1,2,3,4)\bigl |^{2}\, ,
\label{eq:M4t}
\end{equation}
where we again use $s_{ij}=k_i\cdot k_j$ and use the shorthand notation
\begin{equation}
\bigl |s_{12}s_{23}\, \mathcal{A}(1,2,3,4)\bigl |^{2}=\bigl (s_{12}s_{23}\, \mathcal{A}(1,2,3,4)\bigl )\bigl (s_{12}s_{23}\, \widetilde{\mathcal{A}}(1,2,3,4)\bigl )\, .
\end{equation}
Due to \eqref{eq:NSNSvertex} and \eqref{eq:thetapol}, this expression encodes the scattering of any combination of gravitons, antisymmetric tensors and dilatons.\footnote{Worldsheet parity ensures that this expression vanishes for scattering amplitudes involving an odd number of $b_{\mu\nu}$ insertions.}
As in the open string case, the closed-string tree-level and one-loop four-point amplitudes share the same kinematical structure. In particular, we have \cite{Green:1999pv}
\begin{equation}\label{eq:M4l}
    \mathcal{M}^{\mathrm{loop}}_4\sim(\alpha')^3 \, \left(\fft{2\pi^2}{3}+(\alpha')^3\fft{\zeta(2)\zeta(3)}2s_{12}s_{23}s_{13}+\cdots\right)\, K+\cdots\, ,
\end{equation}
where the kinematic structure $K$ is again specified in terms of the $t_8$ tensor \eqref{eq:t8} as
\begin{equation}
    K = t^{\nu_1\ldots\nu_8} t_{\mu_1\ldots\mu_8}\, \theta_1^{[\mu_1}{}_{[\nu_1}k_1^{\mu_2]}k_{1,\nu_2]}\cdots \theta_4^{[\mu_7}{}_{[\nu_7}k_4^{\mu_8]}k_{4,\nu_8]}\, .
\end{equation}

As a final step, we now turn to five-point amplitudes.
We begin with the tree-level closed five-point function which is obtained from the KLT relations.
The tree-level open-string five-point amplitude was originally computed using the covariant approach in \cite{Kitazawa:1987xj,Medina:2002nk,Barreiro:2005hv,Stieberger:2006te} and using the pure spinor formalism \cite{Berkovits:2000fe} in \cite{Mafra:2009bz}.\footnote{For higher-point open string amplitudes from the pure spinor formalism, we refer to \cite{Mafra:2011nv,Mafra:2011nw} for tree-level results and to \cite{Mafra:2012kh,Mafra:2018nla,Mafra:2018pll,Mafra:2018qqe} for loop computations.}
In particular, \cite{Mafra:2009bz} computed the corresponding closed-string tree amplitude using the KLT relations \cite{Kawai:1985xq} which, at the eight-derivative level, takes the form \cite{Schlotterer:2012ny,Boels:2013jua,Green:2013bza}
\begin{equation}\label{eq:M5t}
 \mathcal{M}_5^{\mathrm{tree}}\sim\, (\alpha')^3\, 2\zeta(3)\begin{pmatrix}\widetilde{\mathcal{A}}(1,2,3,5,4)\\
 \widetilde{\mathcal{A}}(1,3,2,5,4)\end{pmatrix}^TS_0M_3\begin{pmatrix}\mathcal{A}(1,2,3,4,5)\\ \mathcal{A}(1,3,2,4,5)\end{pmatrix}\, .
\end{equation}
As before, we denote the Yang-Mills five-point amplitude by $\mathcal{A}(1,2,3,4,5)$ and introduced the $2\times2$ matrices $S_0$ and $M_3$ which take the form
\begin{equation}
S_0=\begin{pmatrix}s_{12}(s_{13}+s_{23})&s_{12}s_{13}\\s_{12}s_{13}&s_{13}(s_{12}+s_{23})\end{pmatrix},\qquad
M_3=\begin{pmatrix}m_{11}&m_{12}\\m_{21}&m_{22}\end{pmatrix}\, ,
\end{equation}
in terms of $s_{ij}=k_i\cdot k_j$ and
\begin{align}
m_{11}&\kern5em=s_3\left(-s_1(s_1+2s_2+s_3)+s_3s_4+s_4^2\right)+s_1s_5(s_1+s_5),\nn\\
m_{12}&\kern5em=-s_{13}s_{24}(s_1+s_2+s_3+s_4+s_5),\nn\\
m_{21}&=m_{12}\big|_{2\leftrightarrow3}=-s_1s_3(s_{13}+s_2+s_{24}+s_4+s_5),\nn\\
m_{22}&=m_{11}\big|_{2\leftrightarrow3}=s_{24}\left(-s_{13}(s_{13}+2s_2+s_{24})+s_{24}s_4+s_4^2\right)+s_{13}s_5(s_{13}+s_5)\, ,
\end{align}
where $s_i\equiv s_{i,i+1}$. Note that, from the ten Mandelstam invariants $s_{ij}$ with $i<j$, only five are independent at the level of massless five-particle kinematics.
We obtain the explicit expression for $\mathcal{A}(1,2,3,4,5)$ from \cite{pss} which contains poles due to factorisation on intermediate gluon states.
The closed string amplitude from the KLT product \eqref{eq:M5t} naively appears to have pole-squared terms, but explicit evaluation shows that these terms actually cancel.
In contrast, there remain single pole terms from intermediate NSNS closed string states which have to be dealt with to construct the five-point contact terms.

The final missing ingredient is the one-loop closed string five-point function.\footnote{Alternatively, the one-loop five-point open-string amplitude was reexamined in \cite{Mafra:2009wi}, and the corresponding closed-string amplitude can, in principle, be derived using chiral splitting and the double-copy construction. Earlier computations of the one-loop amplitude were performed in the Green-Schwarz formalism \cite{Frampton:1985uw,Frampton:1986ea,Frampton:1986gi,Lam:1986kg}, and in the even-even spin structure using the covariant formalism in \cite{Montag:1992dm}.}
A direct computation of the closed-string five-point one-loop amplitude was performed in \cite{Richards:2008jg} using the Green-Schwarz formalism which led to
\begin{align}\label{eq:M5l}
 \mathcal{M}_5^{\mathrm{loop}}&\sim-\fft1{s_{12}}\Bigl[k_1\cdot e_2\,  t_8(e_1,k_1+k_2)-k_2\cdot e_1\,  t_8(e_2,k_1+k_2)-e_1\cdot e_2\,  t_8(k_1,k_2)-s_{12}t_8(e_1,e_2)\Bigr]\nn\\[0.3em]
&\qquad\times\Bigl[k_1\cdot\tilde{e}_2\,  \tilde{t}_8(\tilde{e}_1,k_1+k_2)-k_2\cdot\tilde{e}_1\,  \tilde{t}_8(\tilde{e}_2,k_1+k_2)-\tilde{e}_1\cdot\tilde{e}_2\, \tilde{t}_8(k_1,k_2)-s_{12}\, \tilde{t}_8(\tilde{e}_1,\tilde{e}_2)\Bigr]\nn\\[0.3em]
&\qquad-e_1\cdot\tilde{e}_2 \, t_8(e_2,k_2)\,\tilde{t}_8(\tilde{e}_1,k_1)-e_2\cdot\tilde{e}_1 \, t_8(e_1,k_1)\,\tilde{t}_8(\tilde{e}_2,k_2)\nn\\[0.3em]
&\quad\mp\tfrac{1}{4}\Bigl[s_{12}\, \epsilon_8(e_1,e_2)\tilde{\epsilon}_8(\tilde{e}_1,\tilde{e}_2)+e_1\cdot\tilde{e}_2\, \epsilon_8(e_2,k_2)\tilde{\epsilon}_8(\tilde{e}_1,k_1)+e_2\cdot\tilde{e}_1\, \epsilon_8(e_1,k_1)\tilde{\epsilon}_8(\tilde{e}_2,k_2)\Bigr]\nn\\[0.3em]
&\quad+\mbox{9 more in the other $s_{ij}$ channels}\nn\\[0.3em]
&\quad-e_1\cdot\tilde{e}_1\,  t_8(e_2,k_2)\tilde{t}_8(\tilde{e}_2,k_2)\mp\tfrac{1}{4}e_1\cdot\tilde{e}_1\, \epsilon_8(e_2,k_2)\tilde{\epsilon}_8(\tilde{e}_2,k_2)\nn\\[0.3em]
&\quad+\mbox{4 more for vertices $2,\ldots,5$}\, ,
\end{align}
where we defined
\begin{equation*}
t_8(\zeta_{1},\zeta_{2})=t_8(\zeta_{1},\zeta_{2},e_3,k_3,e_4,k_4,e_5,k_5)\kom \tilde{t}_8(\zeta_{1},\zeta_{2})=t_8(\zeta_{1},\zeta_{2},\tilde{e}_3,k_3,\tilde{e}_4,k_4,\tilde{e}_5,k_5)
\end{equation*}
and similarly for $\epsilon_{8}$ and $\tilde{\epsilon}_{8}$.
In this expression, the upper signs corresponds to the Type IIA string, while the lower signs correspond to Type IIB. 
The final two lines of Eq.~\eqref{eq:M5l} involve trace polarisations, corresponding to external dilaton states. These contributions are crucial when analysing amplitudes that include dilatons. Notably, such terms were omitted in earlier analyses, including \cite{Richards:2008jg,Richards:2008sa}, which focused exclusively on graviton and antisymmetric tensor scattering.
The inclusion of trace polarisations allows us to probe the full set of NSNS-sector couplings, particularly those involving multiple dilatons or mixed dilaton-graviton interactions. These terms form a major focus of Sect.~\ref{sec:NSNSdilaton} and App.~\ref{sec:amptoact}, where we systematically extract and interpret the corresponding effective interactions from the five-point amplitude. Their proper identification is essential for completing the NSNS effective action at the eight-derivative level, especially in light of duality constraints and dimensional reductions relevant for string compactifications.

\subsection{Effective actions for three-forms and gravitons}\label{sec:effact3formgrav}

We now turn to the reconstruction of the Type II effective action in the NSNS sector, based on the previously discussed string amplitudes. While the amplitudes themselves provide essential data, our ultimate goal is to extract the corresponding eight-derivative effective action that captures the low-energy dynamics of massless fields.
The effective action admits a perturbative expansion in powers of $\alpha'$, the inverse string tension, with each order capturing higher-derivative corrections. We write this schematically as
\begin{equation}
\mathcal{L} = \mathcal{L}^{(0)} + (\alpha')^{3} \mathcal{L}^{(3)} + \ldots,
\end{equation}
where $\mathcal{L}^{(0)}$ is the two-derivative leading-order supergravity action, and $\mathcal{L}^{(3)}$ contains the eight-derivative corrections arising at order $(\alpha')^3$. These corrections can themselves be decomposed into contributions from different perturbative sectors
\begin{equation}
\mathcal{L}^{(3)} = \mathcal{L}^{(3)}_{\text{tree}} + \mathcal{L}^{(3)}_{\text{loop}} + \ldots,
\end{equation}
where $\mathcal{L}^{(3)}_{\text{tree}}$ and $\mathcal{L}^{(3)}_{\text{loop}}$ represent the tree-level and one-loop contributions, respectively. Non-perturbative corrections may also contribute, which we identify in the main text e.g. through suitable modular completions, recall the discussion in Sect.~\ref{sec:review}. This decomposition allows us to systematically match the string amplitudes organised by their genus and external states to local operators in the effective action, thus identifying the full structure of the eight-derivative terms consistent with the underlying string dynamics.

We begin with the three-point function computed in Eq.~\eqref{eq:M3}.
Since the amplitude is of order $k^{2}$, the associated terms in the effective action involve two derivatives only.\footnote{For the heterotic string, already the bosonic three point amplitudes includes terms at order $k^{4}$ associated to four derivative terms like Riemann squared terms $R^{2}$.}
More specifically, by expanding the amplitude in terms of the different polarisations \eqref{eq:thetapol}, we find non-vanishing contributions for $h^{3}$, $b^{2}h$, $b^{2}\phi$ and $h\phi^{2}$ corresponding to effective interactions among gravitons, antisymmetric tensors, and dilatons, respectively.
It is straightforward to verify that all these three-point amplitudes are reproduced by the leading-order Einstein-frame action
\begin{equation}
\cL^{(0)}=\sqrt{-g}\, \bigl [R-\ft12\partial\phi^2-\ft1{12}\,\mathrm{e}^{-\phi}H_{3}^2\bigl ]\, ,
\label{eq:EinsAct}
\end{equation}
where $H_3 = \mathrm{d}B$ is the field strength of the antisymmetric two-form.

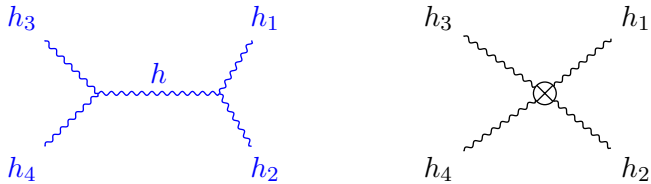
\begin{figure}[t!]
\centering
{\color{blue}
\begin{tikzpicture}[scale=0.8]
\setlength{\feynhanddotsize}{1.ex}
\begin{feynhand}
\vertex (vertex2) at (0.75,-4); 
\vertex (vertex1) at (-1.25,-4); 
\vertex (in1) at (-2.5,-2.75) {$h_{3}$}; 
\vertex (in2) at (-2.5,-5.25) {$h_{4}$}; 
\vertex (out1) at (1.5,-2.75) {$h_{1}$}; 
\vertex (out2) at (1.5,-5.25) {$h_{2}$}; 
\propag [pho] (vertex1) to [edge label = $h$] (vertex2);
\propag [pho] (in1) to (vertex1) ;
\propag [pho] (in2) to (vertex1);
\propag [pho] (vertex2) to (out1) ;
\propag [pho] (vertex2) to (out2);
\end{feynhand}
\end{tikzpicture}
}
\hspace*{1.25cm}
\begin{tikzpicture}[scale=0.8]
\setlength{\feynhandblobsize}{5mm}
\begin{feynhand}
\vertex [crossdot] (vertex2) at (0.75,-4) {}; 
\vertex (in1) at (-1.,-2.75) {$h_{3}$}; 
\vertex (in2) at (-1.,-5.25) {$h_{4}$}; 
\vertex (out1) at (2.25,-2.75) {$h_{1}$}; 
\vertex (out2) at (2.25,-5.25) {$h_{2}$}; 
\propag [pho] (in1) to (vertex2) ;
\propag [pho] (in2) to (vertex2);
\propag [pho] (vertex2) to (out1) ;
\propag [pho] (vertex2) to (out2);
\end{feynhand}
\end{tikzpicture}
\caption{The four-graviton amplitude at closed string tree level, as given in Eq.~\eqref{eq:M4t}, receives contributions from both contact and exchange diagrams. The blue diagram on the left represents the exchange of an intermediate graviton, which gives rise to singular terms in the amplitude. To isolate the four-point contact interaction, one must subtract the full contribution of the diagram on the left, including both its singular and finite parts.}\label{fig:h4}
\end{figure}

We next consider the tree-level four-point function given in Eq.~\eqref{eq:M4t}, which contains two types of contributions. The first one arises from the exchange of intermediate massless NSNS fields using only the vertices from the two-derivative action \eqref{eq:EinsAct}. For example, in the pure graviton case, this corresponds to the diagram shown on the left-hand side of Fig.~\ref{fig:h4}, where a graviton is exchanged between two three-point vertices.
The second contribution is the four-point contact interaction that appears at order $(\alpha')^3$ and cannot be captured by any factorisation through lower-point interactions. To isolate this term, one must subtract the full contribution of the exchange diagrams on the left of Fig.~\ref{fig:h4}, including both their singular and finite parts.
In other words, pole subtraction is not merely a matter of identifying singularities in limits such as $s \rightarrow 0$; rather, it requires computing the full set of field-theory diagrams contributing to each channel and subtracting their total effect.
This subtraction is essential for correctly identifying the higher-derivative structures such as the well-known $t_8 t_8 R^4$ term in the effective action associated with the diagram on the right of Fig.~\ref{fig:h4}.

In fact, the tree-level four-point amplitude $ \mathcal{M}_4^{\mathrm{tree}}$ captures the full NSNS sector of the closed string. 
Looking at the closed-string four-point amplitude at one loop \eqref{eq:M4l}, the leading term directly yields the effective action, as no pole subtraction is needed.\footnote{However, this no longer holds at higher points, where additional contributions must be carefully subtracted just like for the tree level case from above.}
In both cases, the eight-derivative structure arises from the gauge-invariant combination
\begin{equation}
\bar{R}^{\mu_1\mu_2}{}_{\nu_1\nu_2}=2\theta^{[\mu_1}{}_{[\nu_1}k^{\mu_2]}k_{\nu_2]},
\label{eq:Ramp}
\end{equation}
where $\mu_1, \mu_2$ label left-movers and $\nu_1, \nu_2$ label right-movers.
Using the polarisation decomposition in \eqref{eq:thetapol}, this can be rewritten as
\begin{equation}
\bar{R}_{\mu_1\mu_2}{}^{\nu_1\nu_2}=R_{\mu_1\mu_2}{}^{\nu_1\nu_2}+\mathrm{e}^{-\phi/2}\nabla_{[\mu_1}H_{\mu_2]}{}^{\nu_1\nu_2} -\delta_{[\mu_1}{}^{[\nu_1}\nabla_{\mu_2]}\nabla^{\nu_2]}\phi \, ,
\label{eq:Rbar}
\end{equation}
which corresponds to the linearised curvature tensor constructed from a connection with torsion.

Putting everything together, the tree-level and one-loop effective action in Einstein frame reproducing the eight-derivative four-point function then takes the form  \cite{Gross:1986iv,Gross:1986mw}
\begin{equation}\label{eq:Lloop}
(\alpha')^{3}\cL^{(3)} \bigl |_{\text{4-pts}}=\sqrt{-g}\,\alpha \, \biggl [2\zeta(3) \mathrm{e}^{-3\phi/2}+\fft{2\pi^2}{3}\,  \mathrm{e}^{\phi/2}\biggl ] \, t_8t_8 \bar{R}^4+\cdots \, ,
\end{equation}
where we have restored the tree-level numerical factor.  
We note that, although the four-point function is sensitive only to the linearised curvature tensor, the appropriate object in the effective action is the full non-linear curvature tensor $R(\Omega_+)$, as defined below in \eqref{eq:RiemmTWTorsion} This non-linear completion was anticipated in \cite{Kehagias:1997cq}, with further evidence for its structure provided in \cite{Liu:2013dna}. Both of these aspects will manifest explicitly at the level of the five-point function as we will demonstrate below.

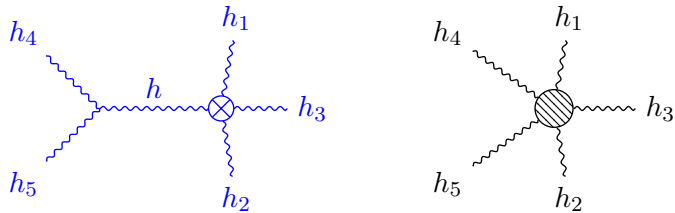
\begin{figure}[t!]
\centering
{\color{blue}
\begin{tikzpicture}[scale=0.8]
\setlength{\feynhanddotsize}{1.ex}
\begin{feynhand}
\vertex [crossdot] (vertex2) at (0.75,-4) {}; 
\vertex (vertex1) at (-1.25,-4); 
\vertex (in1) at (-2.5,-2.75) {$h_{4}$}; 
\vertex (in2) at (-2.5,-5.25) {$h_{5}$}; 
\vertex (out1) at (1.,-2.5) {$h_{1}$}; 
\vertex (out2) at (1.,-5.5) {$h_{2}$}; 
\vertex (out3) at (2.25,-4) {$h_{3}$}; 
\propag [pho] (vertex1) to [edge label = $h$] (vertex2);
\propag [pho] (in1) to (vertex1) ;
\propag [pho] (in2) to (vertex1);
\propag [pho] (vertex2) to (out1) ;
\propag [pho] (vertex2) to (out2);
\propag [pho] (vertex2) to (out3);
\end{feynhand}
\end{tikzpicture}
}
\hspace*{0.75cm}
\begin{tikzpicture}[scale=0.8]
\setlength{\feynhandblobsize}{5mm}
\begin{feynhand}
\vertex [NWblob] (vertex2) at (0.75,-4) {}; 
\vertex (in1) at (-1.,-2.75) {$h_{4}$}; 
\vertex (in2) at (-1.,-5.25) {$h_{5}$}; 
\vertex (out1) at (1.,-2.5) {$h_{1}$}; 
\vertex (out2) at (1.,-5.5) {$h_{2}$}; 
\vertex (out3) at (2.5,-4) {$h_{3}$}; 
\propag [pho] (in1) to (vertex2) ;
\propag [pho] (in2) to (vertex2);
\propag [pho] (vertex2) to (out1) ;
\propag [pho] (vertex2) to (out2);
\propag [pho] (vertex2) to (out3);
\end{feynhand}
\end{tikzpicture}
\caption{The five-graviton amplitude obtained from \eqref{eq:M5t} at closed-string tree level receives contributions from both exchange and contact diagrams. The blue diagram on the left shows the exchange of an intermediate massless graviton $h$ leading to singular terms in the amplitude. To isolate the five-point contact interaction, the full contribution from exchange diagrams like the one on the left must be subtracted, including both their singular and finite parts.}\label{fig:h5}
\end{figure}

Next, we turn to the five-point functions at tree and one-loop.
From the effective action perspective, the five-point amplitudes \eqref{eq:M5t} and \eqref{eq:M5l} receive again two distinct contributions. Similar to the four-point function, the first one arises from intermediate massless NSNS exchange in an $s_{ij}$ channel, but this time connecting a leading-order three-point vertex to a tree-level $(\alpha')^3$ four-point interaction involving the remaining external states. The second contribution comes from five-point contact terms, which encode new higher-derivative couplings beyond those captured in the quartic effective action. These two contributions are illustrated in Fig.~\ref{fig:h5} for the five graviton amplitude, and our focus will be on isolating the latter.

For the tree-level five-point amplitude \eqref{eq:M5t}, the pole subtraction was performed\footnote{We note that the pole subtraction was not done explicitly, but rather inferred via indirect arguments. That is, using the results of \cite{Richards:2008jg} for the one-loop effective action and using the fact that the four-point kinematics is equivalent for tree-level and one-loop, the pole contributions can be removed by simply taking the difference between the tree-level \eqref{eq:M5t} and one-loop expression \eqref{eq:M5l}.} in \cite{Liu:2019ses} and further refined in \cite{Liu:2022bfg} which leads to
\begin{equation}\label{eq:Ltree5pts}
(\alpha')^{3}\cL^{(3)}_{\text{tree}} \bigl |_{\text{5-pts}}=\sqrt{-g}\,\alpha \,  2\zeta(3) \mathrm{e}^{-3\phi/2} \,  \left (\left(t_8t_8-\ft14\epsilon_{8}\epsilon_{8}\right ) R(\Omega_+)^4-4\mathrm{e}^{-\phi} t_{18}\, H_{3}^2R(\Omega_+)^3+\cdots\right ) \, ,
\end{equation}
where we introduce the full non-linear Riemann tensor $R(\Omega_{\pm})$ from a connection $\Omega_{\pm}$ with torsion
\begin{equation}\label{eq:RiemmTWTorsion} 
R(\Omega_{\pm})_{MN}\,^{KL}=R_{MN}\,^{KL}\pm\ee^{-\phi/2}\nabla_{[M} H_{N]}\,^{KL}+\dfrac{\ee^{-\phi}}{2}H_{[M}\,^{K P}H_{N] P}\,^{L}\, .
\end{equation}
Further, we use $\epsilon_{n}\epsilon_{n}$ to denote
\begin{equation}\label{eq:epsilon}
\epsilon_{n\,\mu_1\cdots\mu_{n}}\epsilon_{n}^{\nu_1\cdots\nu_{n}}=n!\delta_{\mu_1}^{[\nu_1}\cdots\delta_{\mu_{n}}^{\nu_{n}]}=-\fft1{m!}\epsilon_{\alpha_1\cdots\alpha_{m}\mu_1\cdots\mu_{n}}\epsilon^{\alpha_1\cdots\alpha_{m}\nu_1\cdots\nu_{n}}\qquad \mbox{with}\quad m+n=10\, ,
\end{equation}
in terms of the generalised Levi-Civita symbol $\epsilon_{n}$ in 10 dimensions.
The higher-dimensional index structure $t_{18}$ has been inferred in \cite{Liu:2022bfg} by duality considerations to M-theory and can e.g.\ be computed using \cite{Peeters:2005tb}. We comment on its role further below.
Upon expanding $R(\Omega_{\pm})$ in \eqref{eq:Ltree5pts} in powers of $H_{3}$, one finds
\begin{equation}
(\alpha')^{3}\cL^{(3)}_{\text{tree}} \bigl |_{\text{5-pts},H_{3}^{2}R^{3}}=\sqrt{-g}\,\alpha \,  2\zeta(3) \mathrm{e}^{-5\phi/2} \left (-4 \, t_{18} +T(\epsilon_{10},t_{8})\right )H_{3}^{2}R^{3}
\end{equation}
in terms of
\begin{equation}\label{eq:KinematicsUniversalG2R3} 
T(\epsilon_{10},t_{8})=-\dfrac{1 }{3}\epsilon_{9}\epsilon_{9}+2\tilde{t}_{8}\tilde{t}_{8}-\dfrac{1}{2}\epsilon_{8}\epsilon_{8}\, .
\end{equation}
Here, the individual contractions of $H_{3}^{2}R^{3}$ are given by
\begin{align}
\label{eq:TTGTRCV2} \tilde{t}_{8}\tilde{t}_{8} H_{3}^{2}R^{3}&=t_{ \mu_{1}\ldots \mu_{8}}t^{ \nu_{1}\ldots \nu_{8}} H^{[ \mu_{1}}\,_{ \nu_{1}\rho}\, H^{ \mu_{2}]\rho}\,_{ \nu_{2}} R^{ \mu_{3} \mu_{4}}\,_{ \nu_{3} \nu_{4}}\ldots R^{ \mu_{7} \mu_{8}}\,_{ N_{7} \nu_{8}}\, ,\\[0.3em]
\label{eq:EEGTRCV2} \epsilon_{8}\epsilon_{8}H_{3}^{2}R^{3}&=\epsilon^{\mu_{1}\ldots \mu_{8}}\epsilon^{\nu_{1}\ldots \nu_{8}} H_{[\mu_{1}| \nu_{1}k}\, H_{|\mu_{2}]}\,^{k}\,_{\nu_{2}}\,  R_{\mu_{3}\mu_{4}\nu_{3}\nu_{4}} \ldots R_{\mu_{7}\mu_{8}\nu_{7}\nu_{8}}\, , \\[0.3em]
\label{eq:EEGTRC} \epsilon_{9}\epsilon_{9}H_{3}^{2}R^{3}&=-\epsilon_{\rho \mu_{0}\ldots \mu_{8}}\epsilon^{\rho \nu_{0}\ldots \nu_{8}}H^{ \mu_{1} \mu_{2}}\,_{ \nu_{0}}\, H_{ \nu_{1} \nu_{2}}\,^{ \mu_{0}}\, R^{ \mu_{3} \mu_{4}}\,_{ \nu_{3} \nu_{4}}\, R^{ \mu_{5} \mu_{6}}\,_{ \nu_{5} \nu_{6}}\, R^{ \mu_{7} \mu_{8}}\,_{ \nu_{7} \nu_{8}}\, .
\end{align}

Turning to the one-loop five-point amplitude \eqref{eq:M5l}, the resulting one-loop effective action has been extensively studied for five-graviton scattering in \cite{Richards:2008jg}, and for mixed amplitudes involving gravitons and antisymmetric tensors in \cite{Peeters:2001ub,Richards:2008sa}. As demonstrated in \cite{Richards:2008jg}, the five-graviton scattering amplitude reproduces the expected one-loop $R^4$ terms in Eq.~\eqref{eq:Lloop} at the non-linear level. In contrast, amplitudes involving both antisymmetric tensors and gravitons generate additional contributions \cite{Richards:2008sa,Liu:2013dna}
\begin{align}
(\alpha')^{3}\cL^{(3)}_{\text{loop}} \bigl |_{\text{5-pts}}&=\sqrt{-g}\, \alpha\, \fft{2\pi^2}{3}\,  \mathrm{e}^{\phi/2}\biggl[\left(t_8t_8\pm\ft14\epsilon_{8}\epsilon_{8}\right ) R(\Omega_+)^4\pm\tfrac{\mathrm{e}^{-\phi}}{3}\epsilon_{9}\epsilon_{9}H_{3}^2R(\Omega_+)^3+\cdots\biggr]\, ,
\label{eq:loopea}
\end{align}
where $\epsilon_{9}\epsilon_{9}H^2R^{3}$ is defined in Eq.~\eqref{eq:EEGTRC}.

Finally, while our focus here has been on the NSNS sector, the complete tree-level quartic effective action, including RR contributions, has been derived using pure spinor techniques \cite{Policastro:2006vt,Policastro:2008hg}.
Even in the absence of additional mixed sector contributions, one can argue that the NSNS sector couplings $H_{3}^{2}R^{3}$ in \eqref{eq:Ltree5pts} and \eqref{eq:loopea} can be uniquely completed in the RR-sector.
As explained in \cite{Liu:2019ses,Liu:2022bfg}, the resulting contributions in the effective can be written in the form
\begin{align}\label{eq:FullResultG2R3TreeLoop} 
\mathcal{L}^{(3)}&\supset \alpha\biggl\{f_{0}(\tau,\bar{\tau})\, \left (T(\epsilon_{10},t_{8})-t_{18}  \right )|G_{3}|^{2}R^{3}+\dfrac{3}{2} \left (f_{1}(\tau,\bar{\tau})\, t_{18}G^{2}_{3}R^{3}+\text{c.c.}\right )\biggl \}
\end{align}
where $T(\epsilon_{10},t_{8})$ is defined in Eq.~\eqref{eq:KinematicsUniversalG2R3}.
The appearance of higher-dimensional index structure $t_{18}$ which does not simply factorise into combinations of $\epsilon_{n}$ or $t_{8}$ can actually be anticipated based on the superspace formalism introduced in \cite{Howe:1983sra}, see also \cite{Peeters:2001ub,deHaro:2002vk,Green:2003an,Green:2005qr,Peeters:2005tb} for related considerations.
Indeed, for the MUV coupling $G_{3}^{2}R^{3}$, the linearised superfield of \cite{Howe:1983sra} uniquely specifies the kinematics in terms of a single superspace integral corresponding to $t_{18}$.

\subsection{Pole subtraction in the scalar-graviton sector}\label{sec:amptoact}

\begin{figure}[t!]
\centering
{\color{blue}
\begin{tikzpicture}[scale=0.6]
\setlength{\feynhanddotsize}{1.ex}
\begin{feynhand}
\vertex [crossdot] (vertex2) at (0.75,-4) {}; 
\vertex (vertex1) at (-1.25,-4); 
\vertex (in1) at (-2.5,-2.75) {$\phi_{1}$}; 
\vertex (in2) at (-2.5,-5.25) {$\phi_{2}$}; 
\vertex (out1) at (1.,-2.5) {$h_{1}$}; 
\vertex (out2) at (1.,-5.5) {$h_{2}$}; 
\vertex (out3) at (2.25,-4) {$h_{3}$}; 
\propag [pho] (vertex1) to [edge label = $h$] (vertex2);
\propag [sca] (in1) to (vertex1) ;
\propag [sca] (in2) to (vertex1);
\propag [pho] (vertex2) to (out1) ;
\propag [pho] (vertex2) to (out2);
\propag [pho] (vertex2) to (out3);
\end{feynhand}
\end{tikzpicture}
\hspace*{0.25cm}
\begin{tikzpicture}[scale=0.6]
\setlength{\feynhanddotsize}{1.ex}
\begin{feynhand}
\vertex [crossdot] (vertex2) at (0.75,-4) {}; 
\vertex (vertex1) at (-1.25,-4); 
\vertex (in1) at (-2.5,-2.75) {$h_{1}$}; 
\vertex (in2) at (-2.5,-5.25) {$\phi_{2}$}; 
\vertex (out1) at (1.,-2.5) {$\phi_{1}$}; 
\vertex (out2) at (1.,-5.5) {$h_{2}$}; 
\vertex (out3) at (2.25,-4) {$h_{3}$}; 
\propag [sca] (vertex1) to [edge label = $\phi$] (vertex2);
\propag [pho] (in1) to (vertex1) ;
\propag [sca] (in2) to (vertex1);
\propag [sca] (vertex2) to (out1) ;
\propag [pho] (vertex2) to (out2);
\propag [pho] (vertex2) to (out3);
\end{feynhand}
\end{tikzpicture}
\hspace*{0.25cm}
\begin{tikzpicture}[scale=0.6]
\setlength{\feynhanddotsize}{1.ex}
\begin{feynhand}
\vertex [crossdot] (vertex2) at (0.75,-4) {}; 
\vertex (vertex1) at (-1.25,-4); 
\vertex (in1) at (-2.5,-2.75) {$h_{1}$}; 
\vertex (in2) at (-2.5,-5.25) {$h_{3}$}; 
\vertex (out1) at (1.,-2.5) {$\phi_{1}$}; 
\vertex (out2) at (1.,-5.5) {$h_{2}$}; 
\vertex (out3) at (2.5,-4) {$\phi_{2}$}; 
\propag [pho] (vertex1) to [edge label = $h$] (vertex2);
\propag [pho] (in1) to (vertex1) ;
\propag [pho] (in2) to (vertex1);
\propag [sca] (vertex2) to (out1) ;
\propag [pho] (vertex2) to (out2);
\propag [sca] (vertex2) to (out3);
\end{feynhand}
\end{tikzpicture}
}
\hspace*{0.25cm}
\begin{tikzpicture}[scale=0.6]
\setlength{\feynhandblobsize}{5mm}
\begin{feynhand}
\vertex [NWblob] (vertex2) at (0.75,-4) {}; 
\vertex (in1) at (-1.,-2.75) {$\phi_{1}$}; 
\vertex (in2) at (-1.,-5.25) {$\phi_{2}$}; 
\vertex (out1) at (1.,-2.5) {$h_{1}$}; 
\vertex (out2) at (1.,-5.5) {$h_{2}$}; 
\vertex (out3) at (2.5,-4) {$h_{3}$}; 
\propag [sca] (in1) to (vertex2) ;
\propag [sca] (in2) to (vertex2);
\propag [pho] (vertex2) to (out1) ;
\propag [pho] (vertex2) to (out2);
\propag [pho] (vertex2) to (out3);
\end{feynhand}
\end{tikzpicture}
\caption{$h^{3}\phi^{2}$ amplitude. Diagrams in blue indicate poles.}\label{fig:h3phi2}
\end{figure}
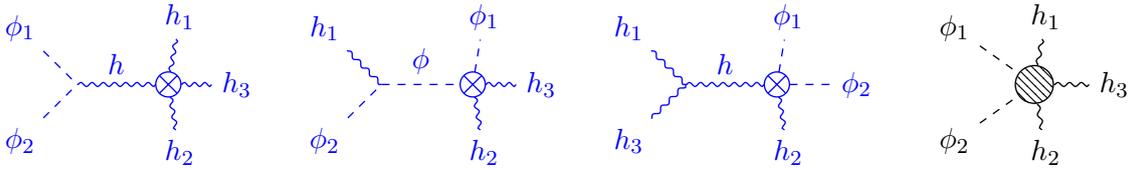

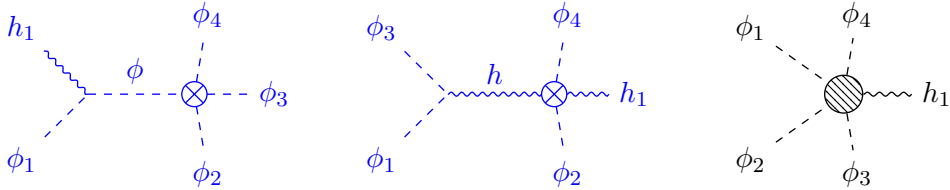
\begin{figure}[t!]
\centering
{\color{blue}
\begin{tikzpicture}[scale=0.7]
\setlength{\feynhanddotsize}{1.ex}
\begin{feynhand}
\vertex [crossdot] (vertex2) at (0.75,-4) {}; 
\vertex (vertex1) at (-1.25,-4); 
\vertex (in1) at (-2.5,-2.75) {$h_{1}$}; 
\vertex (in2) at (-2.5,-5.25) {$\phi_{1}$}; 
\vertex (out1) at (1.,-2.5) {$\phi_{4}$}; 
\vertex (out2) at (1.,-5.5) {$\phi_{2}$}; 
\vertex (out3) at (2.25,-4) {$\phi_{3}$}; 
\propag [sca] (vertex1) to [edge label = $\phi$] (vertex2);
\propag [pho] (in1) to (vertex1) ;
\propag [sca] (in2) to (vertex1);
\propag [sca] (vertex2) to (out1) ;
\propag [sca] (vertex2) to (out2);
\propag [sca] (vertex2) to (out3);
\end{feynhand}
\end{tikzpicture}
\hspace*{0.5cm}
\begin{tikzpicture}[scale=0.7]
\setlength{\feynhanddotsize}{1.ex}
\begin{feynhand}
\vertex [crossdot] (vertex2) at (0.75,-4) {}; 
\vertex (vertex1) at (-1.25,-4); 
\vertex (in1) at (-2.5,-2.75) {$\phi_{3}$}; 
\vertex (in2) at (-2.5,-5.25) {$\phi_{1}$}; 
\vertex (out1) at (1.,-2.5) {$\phi_{4}$}; 
\vertex (out2) at (1.,-5.5) {$\phi_{2}$}; 
\vertex (out3) at (2.25,-4) {$h_1$}; 
\propag [pho] (vertex1) to [edge label = $h$] (vertex2);
\propag [sca] (in1) to (vertex1) ;
\propag [sca] (in2) to (vertex1);
\propag [sca] (vertex2) to (out1) ;
\propag [sca] (vertex2) to (out2);
\propag [pho] (vertex2) to (out3);
\end{feynhand}
\end{tikzpicture}
}
\hspace*{0.5cm}
\begin{tikzpicture}[scale=0.7]
\setlength{\feynhandblobsize}{5mm}
\begin{feynhand}
\vertex [NWblob] (vertex2) at (0.75,-4) {}; 
\vertex (in1) at (-1.,-2.75) {$\phi_{1}$}; 
\vertex (in2) at (-1.,-5.25) {$\phi_{2}$}; 
\vertex (out1) at (1.,-2.5) {$\phi_{4}$}; 
\vertex (out2) at (1.,-5.5) {$\phi_{3}$}; 
\vertex (out3) at (2.5,-4) {$h_{1}$}; 
\propag [sca] (in1) to (vertex2) ;
\propag [sca] (in2) to (vertex2);
\propag [sca] (vertex2) to (out1) ;
\propag [sca] (vertex2) to (out2);
\propag [pho] (vertex2) to (out3);
\end{feynhand}
\end{tikzpicture}
\caption{$h\phi^{4}$ amplitude. Diagrams in blue indicate poles.}\label{fig:h1phi4}
\end{figure}

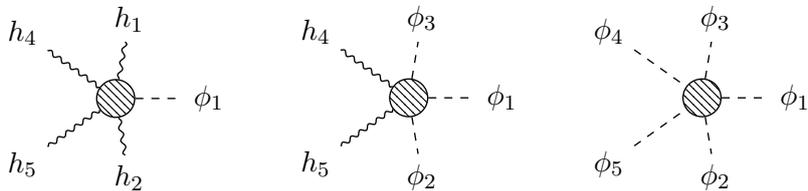
\begin{figure}[t!]
\centering
\begin{tikzpicture}[scale=0.7]
\setlength{\feynhandblobsize}{5mm}
\begin{feynhand}
\vertex [NWblob] (vertex2) at (0.75,-4) {}; 
\vertex (in1) at (-1.,-2.75) {$h_{4}$}; 
\vertex (in2) at (-1.,-5.25) {$h_{5}$}; 
\vertex (out1) at (1.,-2.5) {$h_{1}$}; 
\vertex (out2) at (1.,-5.5) {$h_{2}$}; 
\vertex (out3) at (2.5,-4) {$\phi_1$}; 
\propag [pho] (in1) to (vertex2) ;
\propag [pho] (in2) to (vertex2);
\propag [pho] (vertex2) to (out1) ;
\propag [pho] (vertex2) to (out2);
\propag [sca] (vertex2) to (out3);
\end{feynhand}
\end{tikzpicture}
\hspace*{0.5cm}
\begin{tikzpicture}[scale=0.7]
\setlength{\feynhandblobsize}{5mm}
\begin{feynhand}
\vertex [NWblob] (vertex2) at (0.75,-4) {}; 
\vertex (in1) at (-1.,-2.75) {$h_{4}$}; 
\vertex (in2) at (-1.,-5.25) {$h_{5}$}; 
\vertex (out1) at (1.,-2.5) {$\phi_{3}$}; 
\vertex (out2) at (1.,-5.5) {$\phi_{2}$}; 
\vertex (out3) at (2.5,-4) {$\phi_1$}; 
\propag [pho] (in1) to (vertex2) ;
\propag [pho] (in2) to (vertex2);
\propag [sca] (vertex2) to (out1) ;
\propag [sca] (vertex2) to (out2);
\propag [sca] (vertex2) to (out3);
\end{feynhand}
\end{tikzpicture}
\hspace*{0.5cm}
\begin{tikzpicture}[scale=0.7]
\setlength{\feynhandblobsize}{5mm}
\begin{feynhand}
\vertex [NWblob] (vertex2) at (0.75,-4) {}; 
\vertex (in1) at (-1.,-2.75) {$\phi_{4}$}; 
\vertex (in2) at (-1.,-5.25) {$\phi_{5}$}; 
\vertex (out1) at (1.,-2.5) {$\phi_{3}$}; 
\vertex (out2) at (1.,-5.5) {$\phi_{2}$}; 
\vertex (out3) at (2.5,-4) {$\phi_1$}; 
\propag [sca] (in1) to (vertex2) ;
\propag [sca] (in2) to (vertex2);
\propag [sca] (vertex2) to (out1) ;
\propag [sca] (vertex2) to (out2);
\propag [sca] (vertex2) to (out3);
\end{feynhand}
\end{tikzpicture}
\caption{$h^{4}\phi$, $h^{2}\phi^{3}$, $\phi^{5}$ amplitudes. No pole subtraction necessary.}\label{fig:oddphi}
\end{figure}

We have already analysed the NSNS sector amplitudes and discussed the structure of pole subtractions, particularly for the three-form and graviton couplings, as outlined in Sect.~\ref{sec:effact3formgrav}. We now extend this analysis to the dilaton sector by examining amplitudes involving trace polarisations. All the relevant information is encoded in the known five-point amplitudes; the key task is to carefully identify and subtract the appropriate pole contributions to isolate the five-point contact terms.

Analogous to the five-graviton amplitude $h^5$ shown in Fig.~\ref{fig:h5}, we consider the even dilaton amplitudes $h^3\phi^2$ (Fig.~\ref{fig:h3phi2}) and $h\phi^4$ (Fig.~\ref{fig:h1phi4}). At both tree level and one loop, these amplitudes require pole subtraction to remove contributions from intermediate NSNS exchange diagrams, just as in the pure graviton case.
Since the four-point kinematics for the dilaton in \eqref{eq:Lloop} is universal at tree level and one loop, the pole subtraction procedure at the level of the five-point amplitude is equivalent for both. The results are collected in Table~\ref{tab:FivePointFunction}.

In contrast, for the odd-dilaton amplitudes $h^4\phi$, $h^2\phi^3$, and $\phi^5$ shown in Fig.~\ref{fig:oddphi}, no such subtraction is needed. This is because the effective action contains no odd-dilaton couplings at lower multiplicities, and thus these amplitudes cannot factorise through any lower-point interactions. As a result, their contributions at five points are entirely contact in nature, both at tree level and one loop, see Table~\ref{tab:FivePointFunction}.

As discussed in the main text, obtaining the full completion of the effective action in the RR sector requires input from mixed amplitudes involving RR fields as external states. Unlike the case of $H_3^2 R^3$ couplings in Eq.~\eqref{eq:FullResultG2R3TreeLoop}, pure dilaton-graviton amplitudes are insufficient to fully determine the kinematical structure of the mixed sector. This limitation arises because certain couplings in the RR-completed effective action, particularly those involving more than two dilatons, vanish upon restriction to the pure NSNS sector. This phenomenon was previously observed in \cite{Policastro:2008hg}, where it motivated the introduction of the index structures $\mathcal{O}_1$ and $\mathcal{O}_2$ in Eq.~\eqref{eq:PTaction}. We encounter analogous behaviour in our analysis, as is discussed in Sect.~\ref{sec:RRcontact}.

At tree level, these can be obtained similarly to the pure NSNS sector amplitudes by making use of the KLT relations \eqref{eq:KLT} by tensoring tree-level open-string amplitudes computed in the pure spinor formalism with fermionic external states.
Recall that the pure spinor formalism, originally developed by Berkovits \cite{Berkovits:2000fe} to resolve the limitations of the RNS and GS formalisms, provides a manifestly super-Poincar{\'e} covariant approach to superstring quantisation, see \cite{Mafra:2022wml} for a review.
In particular, this formalism has been used successfully to derive tree-level and one-loop $R^4$ and $D^4 R^4$ couplings \cite{Berkovits:2005ng}, to construct higher-point amplitudes, including five-point tree-level amplitudes with external RR fields \cite{Mafra:2011nv, Mafra:2011nw}, and to match pure spinor results with spacetime supersymmetry constraints from superspace approaches \cite{Green:2019rhz}. For any number of massless external states, the superstring tree-level amplitudes have been computed in \cite{Mafra:2011nv}, see also \cite{Mafra:2022wml} for more compact expressions. In particular, all the tree-level open-string amplitudes required throughout this work have been collected in the database \cite{pss}. The corresponding effective action resulting from these mixed-sector closed-string five-point amplitudes involving at least two RR external states is constructed in Sect.~\ref{sec:RRcomp}.

\begin{figure}[t!]
\centering
{\color{blue}
\begin{tikzpicture}[scale=0.7]
\setlength{\feynhanddotsize}{1.ex}
\begin{feynhand}
\vertex [crossdot] (vertex2) at (0.75,-4) {}; 
\vertex (vertex1) at (-1.25,-4); 
\vertex (in1) at (-2.5,-2.75) {$\chi_1$}; 
\vertex (in2) at (-2.5,-5.25) {$\chi_{2}$}; 
\vertex (out1) at (1.,-2.5) {$\phi_{1}$}; 
\vertex (out2) at (1.,-5.5) {$h_{2}$}; 
\vertex (out3) at (2.25,-4) {$h_{3}$}; 
\propag [sca] (vertex1) to [edge label = $\phi$] (vertex2);
\propag [gho] (in1) to (vertex1) ;
\propag [gho] (in2) to (vertex1);
\propag [sca] (vertex2) to (out1) ;
\propag [pho] (vertex2) to (out2);
\propag [pho] (vertex2) to (out3);
\end{feynhand}
\end{tikzpicture}
\hspace*{0.75cm}
\begin{tikzpicture}[scale=0.7]
\setlength{\feynhanddotsize}{1.ex}
\begin{feynhand}
\vertex [crossdot] (vertex2) at (0.75,-4) {}; 
\vertex (vertex1) at (-1.25,-4); 
\vertex (in1) at (-2.5,-2.75) {$\phi_{1}$}; 
\vertex (in2) at (-2.5,-5.25) {$\chi_{2}$}; 
\vertex (out1) at (1.,-2.5) {$\chi_{1}$}; 
\vertex (out2) at (1.,-5.5) {$h_{2}$}; 
\vertex (out3) at (2.25,-4) {$h_{3}$}; 
\propag [gho] (vertex1) to [edge label = $\chi$] (vertex2);
\propag [sca] (in1) to (vertex1) ;
\propag [gho] (in2) to (vertex1);
\propag [gho] (vertex2) to (out1) ;
\propag [pho] (vertex2) to (out2);
\propag [pho] (vertex2) to (out3);
\end{feynhand}
\end{tikzpicture}
}
\hspace*{0.75cm}
\begin{tikzpicture}[scale=0.7]
\setlength{\feynhanddotsize}{1.ex}
\setlength{\feynhandblobsize}{5mm}
\begin{feynhand}
\vertex [NWblob] (vertex2) at (0.75,-4) {}; 
\vertex (in1) at (-1.,-2.75) {$\chi_{1}$}; 
\vertex (in2) at (-1.,-5.25) {$\chi_{2}$}; 
\vertex (out1) at (1.,-2.5) {$\phi_{1}$}; 
\vertex (out2) at (1.,-5.5) {$h_{2}$}; 
\vertex (out3) at (2.5,-4) {$h_{3}$}; 
\propag [gho] (in1) to (vertex2) ;
\propag [gho] (in2) to (vertex2);
\propag [sca] (vertex2) to (out1) ;
\propag [pho] (vertex2) to (out2);
\propag [pho] (vertex2) to (out3);
\end{feynhand}
\end{tikzpicture}
\caption{$h^{2}\phi\chi^{2}$ amplitude. In contrast to the three dilaton amplitude in Fig.~\ref{fig:oddphi}, there are now exchange channels shown as the diagrams in blue which have to be subtracted to isolate the contact term on the right.}\label{fig:chi2phih2}
\end{figure}
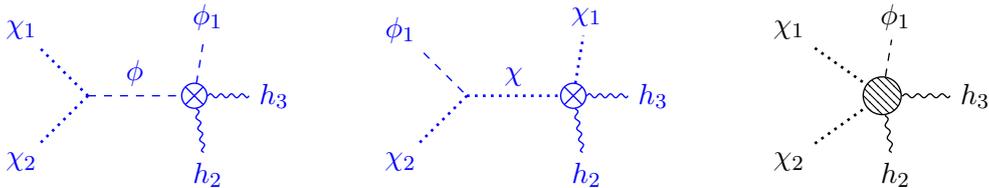

In the context of extracting the effective action from these mixed RR–NSNS sector amplitudes, it is important to note that the pole-subtraction procedure differs significantly from that in the pure NSNS sector. Consider, for instance, the five-point amplitude involving three scalars and two gravitons. In the NSNS sector, no pole subtraction was required for such configurations, as lower-point interactions do not generate couplings with an odd number of dilatons. However, in the mixed sector, this is no longer the case: intermediate exchange diagrams involving lower-order RR–NSNS couplings do contribute. As shown on the left of Fig.~\ref{fig:chi2phih2}, these exchange diagrams must be explicitly subtracted to isolate the five-point contact interaction.

\section{Geometrisation of \texorpdfstring{$\mathrm{SL}(2,\mathbb{Z})$}{SL(2,Z)} through 12D F-theory}\label{app:Ftheory}

Building on the idea that two-derivative axio-dilaton dynamics arise from 12D Einstein gravity via an auxiliary torus, \cite{Minasian:2015bxa} showed that this framework also reproduces the eight-derivative four-point effective action in the scalar–graviton sector. In particular, we write in twelve dimensions\footnote{We note that the original proposal in \cite{Minasian:2015bxa} included a factor of $2\alpha$ in the action \eqref{12EFT}. While this does not pose a problem for the analysis of higher-derivative terms in ten dimensions, it does affect the calculation of the four-dimensional corrections to the Kähler potential. The issue is resolved by accounting for the backreaction arising from solving the twelve-dimensional equations of motion at order $(\alpha')^3$, and subsequently reducing the classical twelve-dimensional Einstein–Hilbert term on this corrected background.}
\begin{align}\label{12EFT}
    \hat{\cL}&=\alpha\, f_{0}(\tau,\bar{\tau})\,\left (t_{8}t_{8}+\dfrac{1}{96}\epsilon_{12}\epsilon_{12}\right )\hat{R}^{4}\, .
\end{align}
Here $\hat{R}$ is the Riemann tensor for a 12D metric $\hat{g}$ simply given by the direct sum of the 10D metric and the 2D $\tau$-dependent metric of a unit-volume two-torus. Note, in particular, the striking fact that the non-standard kinematics referred to as $\mathcal{O}_1$ in \cite{Policastro:2006vt,Policastro:2008hg} (see Eq.~\eqref{eq:PTaction}), which controls the mixed axion/dilaton quartic couplings, lifts to the much more standard structure of $t_8t_8$ in 12D.

It is therefore natural, in our present context, to ask whether this correspondence extends beyond four points, where the proposal of \cite{Minasian:2015bxa} remains untested.  In particular, the terms arising from $\epsilon_{12} \epsilon_{12} \hat{R}^4$ have not yet been matched with any known string amplitude. To address this, we independently carried out the torus reduction of \eqref{12EFT} using complex coordinates, as we will now describe.

\subsection{12D Riemann tensor}

As a first step, we need to compute the 12D Riemann tensor $\hat{R}$ for the metric
\begin{equation}\label{eq:12Dmetric}
g_{MN}=\left (\begin{array}{ccc}
g_{\mu\nu } & 0 & 0 \\ [0.5em]
0 & \tfrac{1}{\tau_{2}} & \tfrac{\tau_{1}}{\tau_{2}} \\ [0.5em]
0 & \tfrac{\tau_{1}}{\tau_{2}} & \tfrac{|\tau|^2}{\tau_{2}}
\end{array} \right )\kom g^{MN}=\left (\begin{array}{ccc}
g^{\mu\nu } & 0 & 0 \\ [0.5em]
0 & \tfrac{|\tau|^2}{\tau_{2}}& -\tfrac{\tau_{1}}{\tau_{2}} \\ [0.5em]
0 & -\tfrac{\tau_{1}}{\tau_{2}} & \tfrac{1}{\tau_{2}} 
\end{array} \right )\kom \tau=\tau_1+\I\,\tau_2\, .
\end{equation}
We write $M,N,\ldots$ for 12D indices, $m,n,\ldots$ for $10$D indices and $\alpha,\beta,\ldots$ for torus indices.
Using the conventions \eqref{eq:tauPG}, we can write for the 10D spacetime derivatives of $\tau_1,\tau_2$
\begin{align}
    \nabla_{\mu}\tau_{1}=-2\tau_{2}Q_{\mu}\kom \nabla_{\mu}\tau_{2}=-\tau_{2}(\cP_{\mu}+\overline{\cP}_{\mu}) \kom Q_{\mu}&=\dfrac{\I}{2}(\cP_{\mu}-\overline{\cP}_{\mu})\, .
\end{align}
We then define
\begin{equation}
    D_{\mu}\cP_{\nu}=\nabla_{\mu}\cP_{\nu}-2\I Q_{\mu}\cP_{\nu}\kom D_{\mu}\overline{\cP}_{\nu}=\nabla_{\mu}\overline{\cP}_{\nu}+2\I Q_{\mu}\overline{\cP}_{\nu}\, .
\end{equation}

To compute the 12D Riemann tensor components $\hat{R}^{M}\,_{NPQ}$, we find it convenient to work with complex coordinates on the torus. The complexified tensor components are obtained from contracting with the appropriate zweibein, namely\footnote{Notice that this differs from \cite{Green:1999by} by exchanging $1\leftrightarrow 2$.}
\begin{equation}
e_{\alpha}^{a}=\dfrac{1}{2\sqrt{\tau_{2}}}\, \left (\begin{array}{cc}
-\tau & 1 \\ 
-\bar{\tau} & 1
\end{array} \right )\kom \alpha=z,\bar{z}\, .
\end{equation}
In this basis,
$T^{2}$-contractions are performed with a constant metric
\begin{equation}
g_{z\bar{z}}=\dfrac{1}{2}\kom g^{z\bar{z}}=2\kom g_{zz}=0\, .
\end{equation}

We then compute
\begin{align}
\hat{R}_{z\bar{z}z\bar{z}}&=-\dfrac{1}{4}|\cP|^{2}\kom \hat{R}_{\mu\nu\, z\bar{z}}=\cP_{[\mu}\overline{\cP}_{\nu]}\kom \hat{R}_{\mu\, z\, \nu\, \bar{z}}=-\dfrac{\cP_{\mu}\overline{\cP}_{\nu}}{2} \kom \hat{R}_{\mu\, z\, \nu\,  z}=\dfrac{D_{\mu}\cP_{\nu}}{2}\, .
\end{align}
Here,
it is interesting to note that, when reducing $\hat{R}^{4}$ to $10$D,
$\hat{R}_{z\bar{z}z\bar{z}}$, $\hat{R}_{\mu\nu\, z\bar{z}}$, and $\hat{R}_{\mu\, z\, \nu\, \bar{z}}$ increase the number of points,
while $\hat{R}_{\mu\, z\, \nu\,  z}$ keeps the number of insertions constant.
That is,
when focussing on four-point amplitudes,
only $\hat{R}_{\mu\, z\, \nu\,  z}+\text{c.c.}$ is relevant as expected from \cite{Policastro:2006vt,Policastro:2008hg}.
It is then straightforward to compute the Ricci tensor and Ricci scalar
\begin{align}
\hat{R}_{z\, \bar{z}}=0\kom \hat{R}_{z\, z}=\dfrac{D_{\mu}\cP^{\mu}}{2} \kom \hat{R}=R-2|\cP|^{2}\, .
\end{align}
Given that we work with a flat torus fibration, $\hat{R}_{z\bar{z}}$ vanishes as expected even without enforcing any 10D equations of motion.

The $12$D vacuum Einstein equations clearly imply the 10D Einstein equations as well as the axio-dilaton equations of motion
\begin{equation}
\hat{R}_{MN}=0\quad\Rightarrow\quad {R}_{\mu\nu}=2\cP_{(\mu}\ov \cP_{\nu)}\kom D_{\mu}\cP^{\mu}=0=D_{\mu}\overline{\cP}^{\mu}
\end{equation}

For reasons that become clear momentarily,
we may also introduce the 12D Weyl tensor
\begin{equation}
\hat{C}_{MNPQ}=\hat{R}_{MNPQ}-{\frac {1}{5}}\left(g_{M[P}\hat{R}_{Q]N}-g_{N[P}\hat{R}_{Q]M}\right)+{\frac {1}{55}}\hat{R}\, g_{M[P}g_{Q]N}\, .
\end{equation}
Let us first look at the components with all legs along the 10D spacetime directions which in terms of the 10D Weyl tensor ${C}_{\mu\nu \rho\sigma}$ read
\begin{align}
\hat{C}_{\mu\nu\rho\sigma}&={C}_{\mu\nu\rho\sigma}-{\frac {1}{5}}\left(g_{\mu[\rho}\left (-\dfrac{1}{4}{R}_{\sigma]\nu}-\cP_{\sigma]}\overline{\cP}_{\nu}-\overline{\cP}_{\sigma]}\cP_{\nu}\right )-g_{\nu[\rho}\left (-\dfrac{1}{4}{R}_{\sigma]\mu}-\cP_{\sigma]}\overline{\cP}_{\mu}-\overline{\cP}_{\sigma]}\cP_{\mu}\right )\right)\nn\\
&\quad+{\frac {1}{55}} \left (-\dfrac{19}{36} R-2|\cP|^{2}\right )\, g_{\mu [\rho}g_{\sigma]\nu}\, .
\end{align}
This implies that if we were to start with an expression only in terms of 12D Weyl tensors,
we would need to apply field redefinitions in 10D to remove any additional Ricci terms to find only 10D Weyl tensors.
If we now apply a field redefinition by setting leftover Ricci terms in the above expressions by dilaton terms,
we obtain
\begin{align}
\hat{C}_{\mu\nu\rho\sigma}&={C}_{\mu\nu\rho\sigma}+{\frac {1}{4}} \left(g_{\mu[\rho}\left (\cP_{\sigma]}\overline{\cP}_{\nu}+\overline{\cP}_{\sigma]}\cP_{\nu}\right )-g_{\nu[\rho}\left (\cP_{\sigma]}\overline{\cP}_{\mu}+\overline{\cP}_{\sigma]}\cP_{\mu}\right )\right)\nn\\
&\quad-\dfrac{1}{18} |\cP|^{2} \, g_{\mu[\rho}g_{\sigma]\nu}\, .
\end{align}

Finally, the remaining Weyl-tensor components with legs on the torus read
\begin{align}
\hat{C}_{\mu\nu \, z \bar{z}}&=\hat{R}_{\mu\nu \, z\bar{z}}\kom\hat{C}_{z \bar{z}\, z \bar{z}}=\hat{R}_{z \bar{z}\, z\bar{z}} -{\frac {\hat{R}}{440}}\, , \nn\\
\hat{C}_{z\nu\, z \sigma}&=\hat{R}_{z\nu z\sigma}- g_{\nu\sigma}\dfrac{\hat{R}_{zz}}{10} \kom\hat{C}_{\bar{z}\nu\, z \sigma}=\hat{R}_{\bar{z}\nu z\sigma}-{\frac {\hat{R}_{\sigma\nu}}{20}} +{\frac {\hat{R} g_{\sigma\nu}}{220}} \, .
\end{align}
Again,
enforcing that the 10D action is expressed only in terms of 10D Weyl tensors (hence no freely appearing Ricci terms),
we may use the 12D equations of motion to set $\hat{R}_{\mu\nu}=\hat{R}=0$ and one finds
\begin{align}
\hat{C}_{\mu\nu \, z \bar{z}}&=\hat{R}_{\mu\nu \, z\bar{z}}\kom \hat{C}_{z \bar{z}\, z \bar{z}}=\hat{R}_{z \bar{z}\, z\bar{z}} \kom \hat{C}_{z\nu z \sigma}=\hat{R}_{z\nu z\sigma}\kom \hat{C}_{\bar{z}\nu z \sigma}=\hat{R}_{\bar{z}\nu z\sigma} \, .
\end{align}

\subsection{Dimensional reduction}

Using the non-vanishing components of the Weyl tensor identified above, we now return to \eqref{12EFT} and perform the explicit dimensional reduction to ten dimensions. Rather than working directly with the twelve-dimensional expression given in \eqref{12EFT}, we employ a shortcut by exploiting the fact that the $R^4$ couplings in ten dimensions can be expressed as two distinct contractions of four Weyl tensors, see, e.g., \cite{Liu:2022bfg} for a detailed discussion. Lifting this structure to twelve dimensions then yields an effective action of the form\\
\begin{equation}\label{12EFTWeyl}
\resizebox{0.9\linewidth}{!}{$%
\hat{\cL}=\dfrac{f_{0}}{16}\,\left (-\tfrac{1}{4}\hat{C}^{MNPQ}\hat{C}_{MN}\,^{RS}\hat{C}_{PR}\,^{TU}\hat{C}_{QSTU}+\hat{C}^{MNPQ}\hat{C}_{M}\,^{R}\,_{P}\,^{S}\hat{C}_{R}\,^{T}\,_{N}\,^{U}\hat{C}_{STQU}\,\right )\, .
$%
} 
\end{equation}
By plugging in the components of the Weyl tensor computed previously, the structure of the contractions makes it clear which types of dilaton couplings can appear. Comparing the resulting terms with those listed in Appendix B of \cite{Minasian:2015bxa}, we find complete agreement.\footnote{We note however that the factor of $1/8$ in Eq.~(B.11) in \cite{Minasian:2015bxa} should only multiply $\epsilon_{10}\epsilon_{10}R^4$ instead of the full bracket.}

Now, to match the result with the five-point function obtained in \eqref{FinalEFT}, we first have to recall that the mere presence of $\mathrm{U}(1)$-violating couplings at five points is a clear obstruction due to the chosen ansatz \eqref{12EFTWeyl} for the 12D couplings. Nonetheless, it might still be possible that the twelve-dimensional action, upon dimensional reduction to ten dimensions, reproduces the couplings in the even scalar sector at five points which is always $\mathrm{U}(1)$ preserving. Restricting ourselves to couplings with two dilatons and three gravitons only, we already find a mismatch between the 10D action \eqref{FinalEFT} and the dimensional reduction of \eqref{12EFT}
\begin{equation}\label{Mismatch}
    \Delta \cL^{(4)} =-8\, \alpha\, f_{0}(\tau,\bar{\tau})\, \epsilon_{8}\epsilon_{8}R^{2}\, (\delta\nabla^2\phi)^{2}+\text{Ricci}\, .
\end{equation}
Upon using 10D equations of motion, the Ricci terms contribute to the four-dilaton function at five points and beyond. Note that the mismatch \eqref{Mismatch}, at the level of the action, is not in conflict with the results of \cite{Minasian:2015bxa}: $\epsilon_{8}\epsilon_{8}R^{2}\, (\delta\nabla^2\phi)^{2}$ starts contributing at the level of the five-point function.

On closer examination, this mismatch should not come as a surprise. If we think of how such 10D couplings are obtained from a Schwinger-loop computation in 11D \cite{Green:1997tv,Green:1997di,Green:1997as,Green:1999by}, we realise that the zero mode reduction performed in \cite{Minasian:2015bxa} can only account for the contribution of amplitudes involving no insertions of KK modes of the 11D superparticle running in the loop. Recall \cite{Green:1999by,Liu:2022bfg} that the $P$-point amplitude with the insertion of $m$ holomorphic and $n$ anti-holomorphic units of the KK loop momentum is proportional to the following quantity:
\begin{equation}
    S(P,m,n)=\int \frac{{\rm d}t}{t}\,\frac{t^P}{t^{9/2}}\,\sum_{\vec{l}}P_z^m P_{\bar{z}}^n e^{-tg^{ab}l_{\alpha}l_b}\,,
\end{equation}
where $t$ denotes the Schwinger proper time parameter, $g$ is the torus metric and $\vec{l}$ are the KK loop momenta. The structure of the vertex operators allows for $m+n\leq 2(P-4)$, whereas the amount of $\mathrm{U}(1)$ violation of the 10D couplings which the above amplitude contributes to, is controlled by the \emph{even} number $|m-n|\leq 2(P-4)$. This means that, at four points, we can only have $S(4,0,0)$, leading to all the 10D invariant couplings correctly captured by the 12D lift. At five points, however, we have four different contributions: $S(5,0,0),S(5,1,1),S(5,2,0),S(5,0,2)$, of which only the first one can be encoded in 12D, at least with the given choice of torus metric \eqref{eq:12Dmetric}. Of the other three contributions, $S(5,2,0)$ and $S(5,0,2)$ are $\mathrm{U}(1)$ violating and are therefore proportional to $f_1$, while $S(5,1,1)$ is $\mathrm{U}(1)$ preserving and therefore should be responsible for the mismatch \eqref{Mismatch}. It would be interesting to better clarify this point, and to possibly find a 12D uplift of the full 10D effective action. We leave this investigation to future work.

%
\bibliographystyle{JHEP}
\bibliography{Literatur}
\end{document}